\documentclass[aps,pra,reprint,superscriptaddress]{revtex4-1} 
\usepackage{graphicx}
\usepackage[colorlinks,citecolor=blue,urlcolor=blue,linkcolor=blue]{hyperref}
\usepackage{siunitx}
\usepackage{amsmath}
\usepackage{braket}
\usepackage{booktabs}
\usepackage{txfonts}
\usepackage[T1]{fontenc}
\begin{document}

\title{Measuring higher-order photon correlations of faint quantum light: a  short review}

\author{K.~Laiho}
\email{kaisa.laiho@dlr.de}
\affiliation{Institute of Quantum Technologies, German Aerospace Center (DLR), Wilhelm-Runge-Str. 10, 89081 Ulm, Germany}

\author{T.~Dirmeier}
\affiliation{Max Planck Institute for the Science of Light, Staudtstr.~2, 91058 Erlangen, Germany}
\affiliation{Friedrich-Alexander-Universit\"at Erlangen-N\"urnberg (FAU), Department of Physics, Staudtstr.~7/B2, 91058 Erlangen, Germany}

\author{M.~Schmidt}
\affiliation{Technische Universit\"at Berlin, Institut f\"ur Festk\"orperphysik, Hardenbergstr.~36, 10623 Berlin, Germany}
\affiliation{Physikalisch-Technische Bundesanstalt (PTB), Abbestr.~2-12, 10587 Berlin, Germany}

\author{S. Reitzenstein}
\affiliation{Technische Universit\"at Berlin, Institut f\"ur Festk\"orperphysik, Hardenbergstr.~36, 10623 Berlin, Germany}

\author{C.~Marquardt} 
\affiliation{Institute of Quantum Technologies, German Aerospace Center (DLR), Wilhelm-Runge-Str. 10, 89081 Ulm, Germany}
\affiliation{Max Planck Institute for the Science of Light, Staudtstr.~2, 91058 Erlangen, Germany}
\affiliation{Friedrich-Alexander-Universit\"at Erlangen-N\"urnberg (FAU), Department of Physics, Staudtstr.~7/B2, 91058 Erlangen, Germany}

\begin{abstract}
Normalized correlation functions provide expedient means for determining the photon-number properties of light. These higher-order moments, also called  the normalized factorial moments of photon number, can be utilized both in the fast state classification and in-depth state characterization.  Further, non-classicality criteria have been derived based on their properties. Luckily, the measurement of the normalized higher-order moments is often loss-independent making their observation with lossy optical setups and imperfect detectors experimentally appealing. The normalized higher-order moments can for example be extracted from the photon-number distribution measured with a true photon-number-resolving detector or accessed directly  via manifold coincidence counting in the spirit of the Hanbury Brown and Twiss experiment. Alternatively, they can be inferred via homodyne detection. Here, we provide an overview of different kind of state classification and characterization tasks that take use of normalized \emph{higher-order} moments and consider different aspects in measuring them with free-traveling light.
\end{abstract}

\maketitle

\section{Introduction}

The definition of light  via discrete energy quanta has in the past brought up a variety of methods for determining its photon-number properties. Nowadays, accessing the photon-number content of light lies in the core of quantum optics experiments. The quantum theory of photon correlations, which  explores the higher-order factorial moments of the photon-number distribution, was seminally formulated in 1963 by Glauber \cite{Glauber1963, Glauber1963a} and Sudarshan \cite{Sudarshan1963}. A few years earlier in 1956 Hanbury Brown and Twiss had conducted their famous experiment, in which they were able to discern between thermal and Poissonian light  \cite{Brown1956a, Brown1956}. Additionally, only shortly after a non-classicality criterium for light was derived in terms of the normalized higher-order moments \cite{Titulaer1965}. 

The normalized higher-order moments offer fast and simple tools for the \emph{quantum state classification} aiming at recognizing what kind of photon-number distribution the investigated light source emits. The most often used classes can directly be accessed via the normalized second-order  moment $g^{(2)}$ and are denoted as the sub-Poissonian $[g^{(2)}<1]$, Poissonian $[g^{(2)} = 1]$ and super-Poissonian $[g^{(2)}>1]$ classes \cite{Fox2006, Migdall2013}. Indeed, they deliver information of the photon-number variance of the emitted light when compared with that of a Poissonian light source having the same mean photon number. 
Another option is to designate the light source directly to a specific class of light, when preferably knowing all its normalized $m$-th order  factorial moments $g^{(m)}$. The most famous classes of quantum light  probably are the thermal light $[g^{(m)}=m!]$, the Poissonian light $[g^{(m)}= 1]$ and the single-photon state $[g^{(m\ge2)}= 0]$ \cite{S.M.Barnett1997}.

The normalized higher-order moments of single-mode light fields incorporating only a single frequency, polarization and wavenumber take constant values \cite{Loudon2000}. Therefore,  it is often convenient to regard the normalized higher-order  moments in their time-independent form. 
However, in general the photon correlations are by definition time dependent and need to be measured in the reference time frame of the emitted photons. Indeed, the time-dependent normalized second-order correlation function $\tilde{g}^{(2)}(\tau)$, in which $\tau$ denotes the relative time difference between the photons emitted from a given light source, has led to another important figure of merit. Namely, photon bunching is observed if $\tilde{g}^{(2)}(\tau = 0) > \tilde{g}^{(2)}(\tau \ne 0)$ and anti-bunching if 
$\tilde{g}^{(2)}(\tau = 0) < \tilde{g}^{(2)}(\tau \ne 0)$ \cite{Gerry2005, Vogel2006}. For photons arriving from two completely independent light fields we expect that $\tilde{g}^{(2)}(\tau) = 1$. In an experiment, the temporal resolution of the used equipment often sets limits to the accuracy of these measurements \cite{Migdall2013}. Therefore, for pulsed light it is usually most convenient to investigate the time-averaged form of the normalized correlation functions.

The measurement of the normalized second-order moment has become a standard tool for verifying and examining the properties of different kind of miniaturized light emitters \cite{Migdall2013}.
Since the first observation of strong anti-bunching from resonance fluorescence of an atom \cite{Kimble1977}, the measurement of $\tilde{g}^{(2)}(\tau = 0)<1$ has become a well-established method for determining the quality of single-photon emitters \cite{Walls1979, Short1983, Leuchs1986, Grangier1986, Diedrich1987, Michler2000}. This has literally resulted to a race towards a value closest to zero, in order to prove that the source emits only a single photon at a time \cite{Meyer-Scott2020, Rodt2020}. Additionally, the measurement of the normalized second-order moment   has become helpful for recognizing the transition to lasing, i.e.~the change from thermal to Poissonian photon-number distribution, in the so-called thresholdless nanolasers that no longer  display a kink in their input-output relation as a sign of ongoing lasing \cite{Jin1994, Khajavikhan2012,Kreinberg2019}. 
Moreover, sources producing photons in pairs, such as the non-linear optical process of parametric down-conversion (PDC) \cite{Burnham1970}, provide an ideal testbed for investigating the photon correlations. The generated twin beams---often called signal and idler---obey strict photon-number correlations \cite{Haderka2005, Avenhaus2008}. The measurement of the joint normalized higher-order moments of signal and idler have proven to be beneficial for accessing the intrinsic properties of the PDC emitters \cite{Razavi2009, Christ2011}. 

Yet, the higher-order moments are also capable  for \emph{in-depth quantum state characterization} \cite{Leuchs2015}. Being connected to a more wider concept of the \emph{moment generating function} they allow one to access other fascinating features of light such as the photon-number parity or even the individual photon-number contributions  \cite{S.M.Barnett1997,Mandel1996}. The photon-number parity directly delivers the value of  the Wigner function at the origin of the phase space \cite{Cahill1969a, Cahill1969} that offers an alternative to the state's density matrix for investigating its properties. Additionally, the non-classicality criteria connected to the Glauber-Sudarshan $P$-representation, which is another quasi-distribution function, can often be formulated in terms of the normalized higher-order moments and have become especially useful when searching for quantum features of light \cite{Agarwal1992,Miranowicz2010}.

In this review, we explore photon correlations and highlight the use of \emph{the normalized moments with orders higher than two} in the quantum state classification and characterization. 
We start by summarizing theoretical aspects of photon correlations in Sec.~\ref{sec:th}, after which we consider experimental methods required for gaining access to them both in discrete (Sec.~\ref{sec:dv}) and continuous (Sec.~\ref{sec:cv}) variables. Thereafter, in Sec.~\ref{sec:states} we review state classification tasks, in which the higher-order moments have proven to been practical. Finally, we regard state characterization tasks varying from the demonstration of non-classicality  in Sec.~\ref{sec:noncl} to the reconstruction of the state's phase-space properties in Sec.~\ref{sec:mg} before summarizing the findings in  Sec.~\ref{sec:concl}. This tutorial review is epecially suited for readers who are already familiar with introductory quantum optics.

\vspace{10pt}

\section{\label{sec:th}  Photon correlations}
The time-dependent normalized $m$-th order correlation function for light can be determined via \cite{Glauber1963a}
\begin{widetext}
\begin{align}
&\tilde{g}^{(m)}(t_{1}, t_{2}= t_{1}+\tau ,\dots t_{m}) 
 = \frac{\braket{\hat{\textbf{a}}^{\dagger}(t_{1})\hat{\textbf{a}}^{\dagger}(t_{2}) \dots  \hat{\textbf{a}}^{\dagger}(t_{m})\hat{\textbf{a}}(t_{m})\dots
\hat{\textbf{a}}(t_{2}) \hat{\textbf{a}}(t_{1}) }}{\braket{\hat{\textbf{a}}^{\dagger}(t_{1})\hat{\textbf{a}}(t_{1})} \braket{\hat{\textbf{a}}^{\dagger}(t_{2})\hat{\textbf{a}}(t_{2})} \dots \braket{\hat{\textbf{a}}^{\dagger}(t_{m})\hat{\textbf{a}}(t_{m})}},
\label{eq_gnt}
\end{align}
\end{widetext}
 in which $\hat{\textbf{a}}^{\dagger}(t_{\sigma})$ [$\hat{\textbf{a}}(t_{\sigma})$] denotes the photon creation [annihilation] operator at time $t_{\sigma}$ ($\sigma= 1,2, \dots m$).
In the case of temporally-resolved measurements one is often most interested in the value of the higher-order correlation at $t_{\sigma} = 0$ ordinarily denoting the event that the photons are emitted simultaneously and delivering information of the underlying photon-number distribution at that time.
The function $\tilde{g}^{(m)}(t_{1}, t_{2}= t_{1}+\tau ,\dots t_{m})$ provides access to the temporal coherence properties of the emitted photons and most importantly, the dynamics of $\tilde{g}^{(2)}(\tau) $  follow the coherence time of the source \cite{Zhang2005, Foertsch2013, Guo2017}.

Besides $\tilde{g}^{(2)}(\tau)$ can be measured with a high temporal resolution if the photo-detection is assisted by the physical phenomenon of the two-photon absorption \cite{Boitier2009}.
In other cases, it is often difficult to measure the time-dependent $\tilde{g}^{(m)}$-function accurately, since the measured outcome is convolved with the detector's response function and thus smeared out, if the  time jitter of the used detectors and of the counting electronics is longer than the temporal coherence of the source \cite{Kreinberg2019, Razavi2009, Bettelli2010, Blauensteiner2009}.
However, sometimes it can be useful to consider time integration over the measured outcome, which mathematically delivers the time integration over the original $\tilde{g}^{(m)}$-function, since the integral over the detector's response function is normalized to unity \cite{Kreinberg2017}. Physically, it means that the total amount of the detection events under the convolution stays the same; just temporal re-binning of the detection events happens due to the jittering. 
This way, the values of the original  $\tilde{g}^{(2)}(\tau = 0)$ function has at least been reconstructed in Refs \cite{Kreinberg2019,Kreinberg2017} for the miniaturized semiconductor light emitters around the lasing transition by taking the advantage of the modified Siegert relation \cite{Blazek2011} and in Ref.~\cite{Ivanova2006} for the squeezed vacuum states prepared in PDC.

In contrast, when investigating pulsed light, one often measures  the photon correlations in time-averaged manner. Thus, one measures all photons arriving at the detector during an integration time, which is chosen to be much longer than the duration of the optical pulses. Therefore, we replace Eq.~(\ref{eq_gnt}) with \cite{Christ2011}
\begin{widetext}
\begin{align}
& {g}^{(m)}  = \frac{\int d t_{1} \int  d t_{2}\dots \int d t_{m} \braket{\hat{\textbf{a}}^{\dagger}(t_{1})\hat{\textbf{a}}^{\dagger}(t_{2}) \dots  \hat{\textbf{a}}^{\dagger}(t_{m})\hat{\textbf{a}}(t_{m})\dots
\hat{\textbf{a}}(t_{2}) \hat{\textbf{a}}(t_{1}) }}{\int d t_{1}  \braket{\hat{\textbf{a}}^{\dagger}(t_{1})\hat{\textbf{a}}(t_{1})}\int d t_{2}  \braket{\hat{\textbf{a}}^{\dagger}(t_{2})\hat{\textbf{a}}(t_{2})} \dots \int d t_{m}  \braket{\hat{\textbf{a}}^{\dagger}(t_{m})\hat{\textbf{a}}(t_{m})}}.
\label{eq_gnint}
\end{align}
\end{widetext}

Extending the integration limits in Eq.~(\ref{eq_gnint}) to $(-\infty, \infty)$, transforming between the time and frequency domains with the Fourier transform, which can be  expressed  in the form $\hat{\textbf{a}}(t) = \frac{1}{\sqrt{2\pi}} \int d\omega \ \hat{\textbf{a}}(\omega)\textrm{e}^{-\imath \omega t}$ with $\omega$ denoting the angular optical frequency, and utilizing the definition $\delta(\omega) = \frac{1}{2\pi} \int dt_{\sigma} \textrm{e}^{\imath \omega t_{\sigma}}$  \ $(\sigma= 1,\dots m)$ of the delta function, we find
\begin{widetext}
\begin{align}
& {g}^{(m)} = \frac{\int d \omega_{1} d \omega_{2}\dots d \omega_{m} \ \braket{\hat{\textbf{a}}^{\dagger}(\omega_{1})\hat{\textbf{a}}^{\dagger}(\omega_{2}) \dots  \hat{\textbf{a}}^{\dagger}(\omega_{m})\hat{\textbf{a}}(\omega_{m})\dots
\hat{\textbf{a}}(\omega_{2}) \hat{\textbf{a}}(\omega_{1}) }}{\int d \omega_{1} \braket{\hat{\textbf{a}}^{\dagger}(\omega_{1})\hat{\textbf{a}}(\omega_{1})} \int d \omega_{2} \braket{\hat{\textbf{a}}^{\dagger}(\omega_{2})\hat{\textbf{a}}(\omega_{2})} \dots  \int d\omega_{m}\braket{\hat{\textbf{a}}^{\dagger}(\omega_{m})\hat{\textbf{a}}(\omega_{m})}}.
\label{eq_gnomega}
\end{align}
\end{widetext}
Eq.~(\ref{eq_gnomega}) delivers the value for the time-averaged normalized $m$-th order correlation function in the frequency domain.

In the framework of broadband and orthonormal spectral modes introduced in Ref.~\cite{Rohde2007} we define the broadband photon creation operators $\hat{\textbf{A}}^{\dagger}_{j} = \int d\omega \ \varphi_{j}(\omega) \ \hat{\textbf{a}}^{\dagger}(\omega)$ with $\varphi_{j}(\omega)$ denoting the profile of the $j$-th mode, for which it holds 
\begin{align}
\int d \omega \  \varphi^{\star}_{j}(\omega) \ \varphi_{j^{\prime}}(\omega) &= 
\left \{ \begin{array}{l} 1 \quad \textrm{if} \quad j = j^{\prime}\\
                                        0 \quad \textrm{otherwise}\\
                                      \end{array} \right. \quad \textrm{and}
 \nonumber\\ 
 \sum_{j} \varphi_{j}(\omega) \ \varphi^{\star}_{j}(\tilde{\omega}) &= \delta(\omega-\tilde{\omega}).
\label{eq:basis}
\end{align}
Thus, we can re-write Eq.~(\ref{eq_gnomega}) with the help of  the relation $\hat{\textbf{a}}^{\dagger}(\omega) = \sum_{j}\hat{\textbf{A}}^{\dagger}_{j} \ \varphi^{\star}_{j}(\omega)$. This yields
\begin{align}
& {g}^{(m)}   =  \frac{\braket{ : \bigg(  \sum_{j} \hat{\textbf{A}}^{\dagger}_{j} \hat{\textbf{A}}_{j} \bigg )^{m}:}}{\braket{\sum_{j} \hat{\textbf{A}}^{\dagger}_{j}\hat{\textbf{A}}_{j}}^{m}},
\label{eq:q_N}
\end{align}
in which : :  denotes the normal operator ordering. Further, Eq.~(\ref{eq:q_N}) reduces to the well-known definition of the normalized factorial moments of photon number given by
\begin{align}
& {g}^{(m)} = \frac{\braket{: \hat{N}^{m}:}}{\braket{\hat{N}}^{m}},
\label{eq:gnN}
\end{align}
when written in terms of the multi-mode photon-number operator $\hat{N} = \sum_{j}\hat{\textbf{A}}^{\dagger}_{j}\hat{\textbf{A}}_{j}$ that is given as a sum of the photon-number operators of the independent and thus commuting modes.

A more suitable form for evaluating the normalized higher-order factorial moments is available, if one has an access to the photon-number content of the investigated light. Thence, they can be expressed as \cite{S.M.Barnett1997}
 \begin{align}
\varg^{(m)} &= \frac{\sum_{n}n(n-1)\dots(n-m+1)\rho_{n}}{\left (\sum_{n} n\rho_{n} \right )^{m}}
= \frac{\sum_{n} \frac{n!}{(n-m)!}\ \rho_{n}}{\left( \sum_{n} n  \rho_{n}\right )^{m}},
\label{eq:gm_stat}
\end{align} 
in which $\rho_{n}$ denotes the photon-number distribution of the investigated quantum optical state, or shortly its photon statistics, given in terms of the photon number $n$.
Interestingly, Eqs~(\ref{eq:gnN}) and (\ref{eq:gm_stat}) are equivalent, as can be seen, if the photon-number distribution of the investigated state is re-written as $\rho_{n} = \braket{: \textrm{e}^{-\hat{N}}\frac{\hat{N}^{n}}{n!}:}$ and this expectation value is replaced in Eq.~(\ref{eq:gm_stat}). By using $\textrm{e}^{\hat{N}} = \sum_{n}\frac{\hat{N}^{n}}{n!}$ and by utilizing Baker-Hausdorff formula for commuting operators we gain
\begin{align}
\varg^{(m)} & = \frac{\braket{: \textrm{e}^{-\hat{N}}\frac{\sum_{n}\hat{N}^{n-m}}{(n-m)!}\hat{N}^{m}:}}{ \braket{: \textrm{e}^{-\hat{N}}\frac{ \sum_{n} \ \hat{N}^{n-1}}{(n-1)!} \hat{N}:}^{m}} =  \frac{\braket{: \hat{N}^{m}:}}{\braket{\hat{N}}^{m}}.
\label{eq:gnP}
\end{align}
As suggested by our treatment, the normalized higher-order moments can be evaluated either from the manifold coincidence discrimination or via the photon statistics measured with a true photon-number-resolving detector. Next, we shortly review important  aspects of both these measurement schemes in discrete variables.

\section{\label{sec:dv}Measurement of the normalized higher-order moments in discrete variables}

The normalized higher-order moments have the potential of becoming important in quantum optical metrology \cite{Moreva2019, Rebufello2019}. However, experimental imperfections can introduce artifacts in the observed values \cite{Stensson2018, Guo2020} and therefore a careful analysis of the measurement arrangement is necessary \cite{Chesi2019} in order to validate the initial assumptions made. We start by investigating the Hanbury Brown and Twiss (HBT) experiment that can resolve the normalized second-order moment, after which we extend the beam splitter network to be able to measure moments of orders higher than two. Thereafter, we review methods revealing the normalized higher-order moments from the photon statistics.

\subsection{\label{sec:hbt}Two-fold coincidence discrimination}
\begin{figure}[!tb]
\centering \includegraphics[width = 0.40\textwidth]{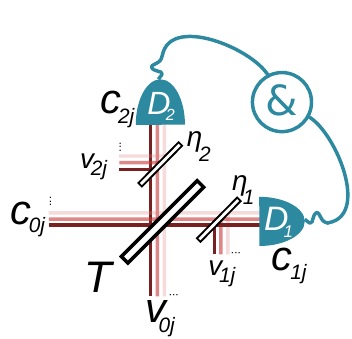}
\caption{\label{fig:HBT} Experimental arrangement for an HBT measurement realized by sending the investigated light beam with $j$ independent modes to a beamsplitter having a splitting ratio of $T$. Optical losses of the detectors are modeled with beam splitters having splitting ratios of $\eta_{1}$ and $\eta_{2}$ in front of the lossless detectors $D_{1}$ and $D_{2}$, respectively. For more details see text.}
\end{figure}

In a typical HBT experiment, as sketched in Fig.~\ref{fig:HBT}, the investigated light beam is sent through a beam splitter with an optical vacuum filling its other input port, while  coincidence counts between the two detectors  placed at its outputs as well as the single counts at each one of them are recorded. The normalized second-order  moment is then typically evaluated via
\begin{align}
{g}^{(2)}  = \frac{\braket{\hat{N}_{1}\hat{N}_{2}}}{\braket{\hat{N}_{1}}\braket{\hat{N}_{2}}} \approx \frac{P_{c}}{P_{D_{1}}P_{D_{2}}},
\label{eq:g2HBT}
\end{align}
in which   $\braket{\hat{N}_{1}\hat{N}_{2}}$ denotes the expectation value for the coincidences and  $\braket{\hat{N}_{1}}$ and $\braket{\hat{N}_{2}}$ are  the average photon numbers in the two output arms. 
We replace the expectation values on the right-hand side in Eq.~(\ref{eq:g2HBT}) with $P_{c}$ denoting the probability of a coincidence click  measured between the  detectors $D_{1}$  and $D_{2}$ and $P_{D_{1}}$ and $P_{D_{2}}$ representing  the single click probabilities at these detectors, correspondingly.
This approximation is possible, only when the detectors' responses are linear with respect to the photon number. For click detectors---sometimes also called bucket detectors---that are able to resolve only between the optical vacuum and at least one photon impinging on such a detector, this requirement can only be satisfied for faint light, that is, highly attenuated light beams \cite{click_det}.

We estimate the expectation values in Eq.~(\ref{eq:g2HBT}) for the scheme in Fig.~\ref{fig:HBT} by assuming that all optical modes experience  the same  beam splitter ratio and the same amount of losses in a given beam splitter output arm. The mean photon numbers at the beam splitter input and output arms (labelled as $\varsigma = 0,1,2$ in Fig.~\ref{fig:HBT}) are evaluated via $\hat{N}_{\varsigma} =  \sum_{j}\hat{c}^{\dagger}_{\varsigma j}\hat{c}_{\varsigma j}$, in which $j$ denotes the independent input modes, for which it applies $[\hat{c}_{\varsigma j},\hat{c}^{\dagger}_{\varsigma^{\prime} j^{\prime}}] = \delta_{j = j^{\prime}}\delta_{\varsigma = \varsigma^{\prime}}$ with $\delta$ being the Kroenecker delta function. By using the beam splitter transformations
\begin{align}
\hat{c}_{1j}& \rightarrow \sqrt{\eta_{1}}\left ( \sqrt{T}\hat{c}_{0j}+\sqrt{1-T}\hat{v}_{0j}\right ) + \sqrt{1-\eta_{1}}\hat{v}_{1j} \quad \textrm{and}\nonumber \\
\hat{c}_{2j}&\rightarrow \sqrt{\eta_{2}}\left ( \sqrt{T}\hat{v}_{0j}-\sqrt{1-T}\hat{c}_{0j}\right ) + \sqrt{1-\eta_{2}}\hat{v}_{2j},
\end{align}
where $\hat{v}_{ij}$ ($i = 0,1,2$) denotes the ports filled with optical vacuum, we re-write the expectation values in Eq.~(\ref{eq:g2HBT}) in terms of the photon creation and annihilation operators at the beam splitter input port and trace over the optical vacuum modes.
The average mean photon numbers deliver $\braket{\hat{N}_{1}}= \eta_{1}T\braket{\sum_{j}\hat{c}^{\dagger}_{0j}\hat{c}_{0j}} = \eta_{1}T\braket{\hat{N}_{0}}$ and $\braket{\hat{N}_{2}}= \eta_{2}(1-T)\braket{\sum_{j}\hat{c}^{\dagger}_{0j}\hat{c}_{0j}} = \eta_{2}(1-T) \braket{\hat{N}_{0}}$.
Due to the commuting modes at the beam splitter output $\braket{\hat{N}_{1}\hat{N}_{2}} = \braket{\sum_{ j,j^{\prime}}\hat{c}^{\dagger}_{1j} \hat{c}_{1j} \hat{c}^{\dagger}_{2j^{\prime}}\hat{c}_{2j^{\prime}} } = \braket{\sum_{ j,j^{\prime}}\hat{c}^{\dagger}_{1j}  \hat{c}^{\dagger}_{2j^{\prime}}\hat{c}_{2j^{\prime}}\hat{c}_{1j} }= \eta_{1}\eta_{2}T(1-T)\braket{\sum_{ j,j^{\prime}}\hat{c}^{\dagger}_{0j} \hat{c}^{\dagger}_{0j^{\prime}}\hat{c}_{0j^{\prime}}\hat{c}_{0j} } = \eta_{1}\eta_{2}T(1-T) \braket{:\hat{N}_{0}^{2}:}$.
Finally, we find
\begin{align}
{g}^{(2)} &=  \frac{\braket{\hat{N}_{1}\hat{N}_{2}}}{\braket{\hat{N}_{1}}\braket{\hat{N}_{2}}}  = \frac{\braket{:\hat{N}^{2}_{0}:}}{\braket{\hat{N}_{0}}^{2}}.
\label{eq:g2_mm}
\end{align}
As indicated in Eq.~(\ref{eq:g2_mm}), the evaluation of the normalized second-order moment in the HBT arrangement is in most cases independent of the beam splitter ratio and optical losses of the employed detectors. This is often justified, since the optical components are  typically selected such that they exhibit spectrally only slowly varying characteristics over the bandwidth of the investigated light emitters.
However, more sophisticated treatment is required, if the different modes suffer from different amount of optical losses, like for example in the case of spectrally multimodal PDC emission, which is filtered to a desired bandwidth \cite{Tapster1998, Branczyk2010}. 

\subsection{Manifold coincidence discrimination}
\begin{figure}[!tb]
\centering
\includegraphics[width = 0.45\textwidth]{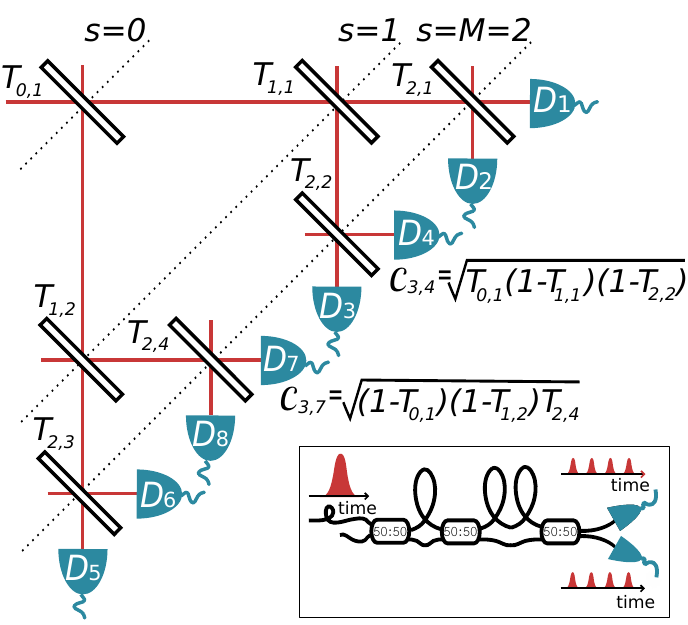}
\caption{\label{fig:exHBT} A sketch of a spatially extended beam splitter network for extracting the values of the  normalized higher-order factorial moments via m-fold coincidence counting at the $m$ chosen detectors.  
The inset shows the more often used implementation for pulsed light via time-multiplexed detection. For more details see text.}
\end{figure}

The normalized moments with orders higher than two can be accessed by extending the HBT setup to a larger beam splitter network. Such a network can be constructed for example via spatial multiplexing as illustrated in Fig.~\ref{fig:exHBT}. 
The evaluation of the normalized $m$-th order moment can be carried out  in analog to the HBT experiment  by counting manifold coincidences. One then arrives at \cite{Avenhaus2010}
\begin{align}
{g}^{(m)} &= \frac{\braket{\hat{N}_{1}\hat{N}_{2}\dots \hat{N}_{m}}}{\braket{\hat{N}_{1}}\braket{\hat{N}_{2}}\dots \braket{\hat{N}_{m}}}
 = \frac{\braket{:\hat{N}_{0}^{m}:}}{\braket{\hat{N}_{0}}^{m} },
\label{eq:gmfold}
\end{align}
in which $\hat{N}_{o}$ ($o = 1,2,\dots m$) denotes the mean-photon number operator at the output ports of the beam splitter network and 
$\hat{N}_{0}$ that at the input port.

We exemplify this method on the basis of the beam splitter network in Fig.~\ref{fig:exHBT} by following Ref.~\cite{Avenhaus2010}.
With  $M$ labeling the depth of the beam splitter network, we denote the stages of the beam splitter network with $s = \{0,\dots M\}$. Thus, the total number of beam splitters at a given stage is $l = \{1,\dots 2^{s}\}$.
Altogether, there are $2l$ outputs at each stage and at the ouput stage they are labeled with $o = \{ 1,\dots 2^{M+1}\}$.  Thus, the highest measurable moment of the network is $m = 2^{M+1}$. At each stage, we denote the beam splitters with their transmission  coefficient $T_{s,l}$ and the vacuum modes with  $\hat{v}_{s,l}$ (not shown in Fig.~\ref{fig:exHBT}).
The beam splitter input-output relations can be expressed with 
\begin{align}
\hat{c}_{s, 2l-1} &= \sqrt{T_{s-1,l}}\ \hat{c}_{s-1,l}+\sqrt{1-T_{s-1,l}}\ \hat{v}_{s-1,l} \quad \textrm{and}\nonumber \\
\hat{c}_{s, 2l} &= -\sqrt{1-T_{s-1,l}}\ \hat{c}_{s-1,l}+\sqrt{T_{s-1,l}}\ \hat{v}_{s-1,l},
\label{eq:network}
\end{align}
 in which the odd values $(2l-1)$ denote the transmission and even values $(2l)$ reflection at a beam splitter. 
In order to evaluate Eq.~(\ref{eq:gmfold}) we require the photon-number operators $\hat{N}_{o} = \hat{c}^{\dagger}_{M+1,o}\hat{c}^{\phantom{\dagger}}_{M+1,o}$ at the output stage $M+1$.
When evaluating the expectation values in Eq.~(\ref{eq:gmfold}), similarly to our treatment in Sec.~\ref{sec:hbt} one recognizes that after tracing over the vacuum modes the photon  creation and annihilation operators  are linearly proportional to those at the input of the network, that is, $\hat{c}^{\dagger}_{s = M+1, o} \propto \mathcal{C}^{\star}_{M+1, o}\hat{c}^{\dagger}_{0,1}$ and $\hat{c}_{s = M+1, o} \propto \mathcal{C}_{M+1, o}\hat{c}_{0,1}$, respectively.
Each path from the input of the beamsplitter network to its ouputs has its individual prefactor, and we denote the ones for $\mathcal{C}_{3, 4}$ and $\mathcal{C}_{3, 7}$ in Fig.~\ref{fig:exHBT} as an example. Clearly, they cancel out when evaluating Eq.~(\ref{eq:gmfold}).

In an experiment, Eq.~(\ref{eq:gmfold}) is approximated by dividing an $m$-fold coincidence probability between $m$ chosen detectors by the product the probabilities of detecting a single click in each of the chosen $m$ detectors.
Similar to the HBT setup, it is also here necessary to approximate the used detectors as linear detectors, for which $P_{D_{i}}  \propto  \eta \braket{\hat{N}_{i}}$, such that the detected count rates are linear with respect to the mean photon number of the light impinging on the detector $D_{i}$. 
Since for click detectors this can be achieved only at low mean photon numbers, the incident beam can be attenuated prior to measurement, if necessary. 
Regarding the network in Fig.~\ref{fig:exHBT}, as the used detectors are binary detectors, there are 256 different measurement outcomes ranging from a no-click event to an 8-fold coincidence-click event.
Such extended HBT measurements have been implemented via spatially or temporally multiplexing beam-splitter networks connected to avalanche-photo diodes or photo-multiplier tubes \cite{Avenhaus2010, Jezek2011, Oppel2012, Qi2018}, interleaved click detectors made of superconducting nanowires \cite{Stevens2010, Goltsman2001, Kitaygorsky2009} and with the help of fast multipixel cameras \cite{Assmann2009, Wiersig2009,Gorobtsov2017}.

\subsection{Measurement of photon statistics}

One can access the normalized higher-order moments also directly from the measured photon-number distribution by utilizing Eq.~(\ref{eq:gm_stat}). The well-known photo-multiplier tubes are capable of measuring the photon statistics \cite{Hong1986, Ramilli2010}, although they  typically suffer from rather low detection efficiencies. In the past, the photon statistics of continuous wave light has also been measured in good approximation with avalanche photo diodes being click detectors \cite{Banaszek1999}. More recently, 
both the most popularly used click detectors---the avalanche photo diodes and the superconducting nanowire detectors---
have also been shown to have an intrinsic photon-number resolution, however being currently only able to count a small number of photons \cite{Thomas2010, Cahall2017}. 

Quasi photon-number-resolving detectors based on spatial or temporal multiplexing and click detection also give access to the photon statistics after considering the probabilistic splitting of photons to different spatial or temporal bins \cite{Achilles2004, Afek2009}.  The effect of this convolution procedure can be formulated in an ensemble measurement in a matrix form.  Apart from that the loss-tolerance can also  be achieved by modeling the losses with an invertable loss-matrix.
This method has been applied for revealing the loss-inverted photon statistics from the recorded loss-degraded click statistics \cite{Avenhaus2008, Achilles2005}. Also other methods such as different kind of optimization routines can be used to access the photon statistics from the measured clicks \cite{Allevi2009, Harder2014, Perina2018}.
Regarding the evaluation of the normalized higher-order moments with Eq.~(\ref{eq:gm_stat}), much care has to be taken that it is evaluated from the lossy photon statistics, i.e. from the backconvoluted one, and not directly from the measured raw click statistics. If the measured raw click statistics, which is lossy and convoluted, is used for evaluating the normalized higher-order moments via  Eq.~(\ref{eq:gm_stat}) corrections on that equation need to be applied \cite{Sperling2012, Bartley2013}.

Better adapted for the measurement of photon statistics are the \emph{true} non-commercial photon-number-resolving detectors, like the visible-light-photon-counter VLPC \cite{Waks2004, Waks2006} and  the transition-edge sensor (TES) \cite{Rosenberg2005, Brida2012}.
Then, the evaluation of the normalized higher-order moments via photon statistics with Eq.~(\ref{eq:gm_stat}) is a straightforward task. However, the accuracy, at which the $m$-th order moment can be extracted, depends both on the underlying photon statistics and also on the amount of statistics taken. 
We exemplify this by evaluating the higher-order moments for Poissonian light measured with a TES. Operating between the superconducting  phase and the normal resistive phase close to its critical temperature, the number of the  photons absorbed in a TES can be read out since the heat from the absorption of a photon causes a gradual transition in the temperature of the TES towards the normal resistive phase. 
Recently, many experiments have taken the advantage of their close-to-unity efficiency \cite{Lita2008}, excellent energy resolution \cite{Schmidt2018} and the capability to count tens of photons \cite{Schmidt2021} for reconstruction photon statistics with TESs. Currently, on the down-side is their rather long temporal resolution of  typically $>\SI{1}{\micro\second}$.

\begin{figure*}[!t]
\centering \includegraphics[width = 0.99\textwidth]{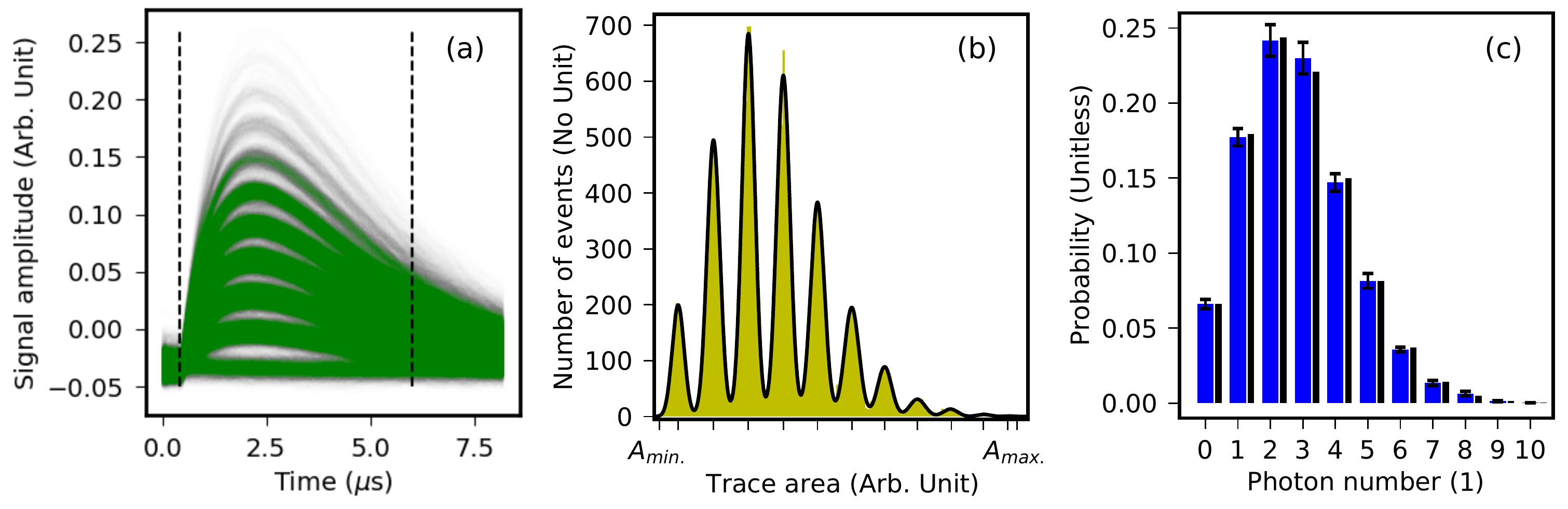}\\
\caption{\label{fig:TES}  (a-c) Reconstruction of the photon statistics for a telecom laser with \SI{1542}{\nano\meter} wavelength  from $2\cdot10^{4}$ TES's traces  (shown in a) measured at \SI{77.5}{\kilo\hertz}. The black solid line in (b), which represents function in the form $\sum_{n} \gamma_{n}\exp{[-\alpha_{n}(A-\beta_{n})^{2}]}$ with $\alpha_{n}$, $\beta_{n}$ and $\gamma_{n}$ being fitting parameters for each photon number $n$ and given in terms of the trace area $A$, is fitted to the histogram of TES's trace areas (yellow bars). The individual photon number contributions illustrated with blue bars in (c) are extracted from the histogram of the trace areas via $\frac{A_{n}}{\sum_{A_{n}}}$ with  $A_{n}$ denoting the area under the peak allocated with the photon number $n$, which can be extracted  for example  via $A_{n} = \gamma_{n}\sqrt{\pi/\alpha_{n}}$. The dashed lines in (a) show the temporal acceptance window for the calculation of the trace areas. The black bars in (c) illustrate a Poissonian distribution with approximately the same mean photon number as the measured photon statistics has.}
\end{figure*}

The area of an individual TES trace is proportional to the measured photon number \cite{Schmidt2018, Schmidt2021}. In Fig.~\ref{fig:TES}(a) we illustrate  typical TES responses in an ensemble measurement of a coherent laser source. The photon statistics can be reconstructed by forming a histogram of the trace areas in Fig.~\ref{fig:TES}(b), and by allocating a photon number for each peak in it. The energy resolution of the detector is determined as the ratio of the full-width-half-maximum of a single histogram peak ($\Delta E$) to the gap between two histogram peaks corresponding to the energy of a quantum ($E = \hbar\omega$). For the used detector we achieve the ratio $\frac{\Delta E}{E} \approx 0.4$ at the single-photon level. Further, the used detector has an efficiency $>70\%$.
In Fig.~\ref{fig:TES}(c) we illustrate the extracted photon statistics (blue bars) and compare it with a Poissonian photon-number distribution (black bars) having the same mean photon number of approximately $2.7$ as the experimentally reconstructed photon statistics has. 
 
\begin{figure}[t]
\centering \includegraphics[width = 0.45\textwidth]{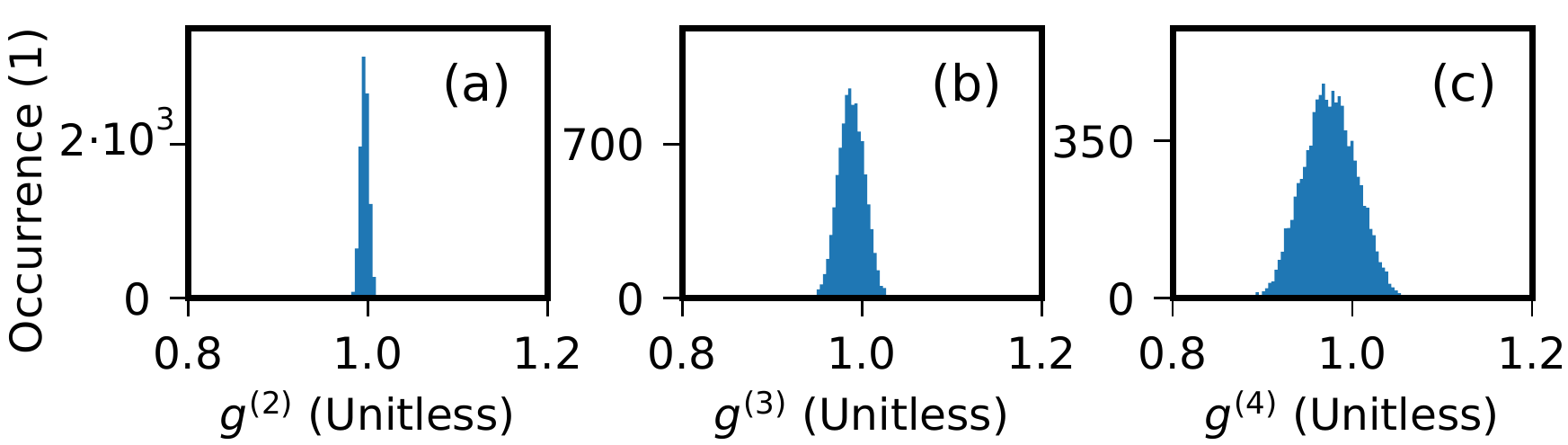}
\caption{\label{fig:TES2} Distributions for the accuracy of the normalized factorial moments extracted from experimental photon statistics in Fig.~\ref{fig:TES}(c) and delivering the values (a) $g^{(2)}= 0.996(5)$, (b)  $g^{(3)}= 0.988(15)$ and (c) $g^{(4)} = 0.975(30)$. The errors correspond to the standard deviation within the simulated ensemble.}
\end{figure}

Now, applying Eq.~(\ref{eq:gm_stat}) we can extract the normalized moments up to the fourth order. 
In order to extract the accuracy, at which these higher-order moments are retrieved from the measured photon statistics, we make a Monte-Carlo simulation with maximally $10^{4}$ trials. 
For this purpose we let each photon-number contribution in Fig.~\ref{fig:TES}(c) vary within its extracted error and calculate the value of Eq.~(\ref{eq:gm_stat}) for each in the ensemble simulated photon statistics.
This way we  gain the accuracy, at which  the normalized higher-order moments are recorded, as shown in Fig.~\ref{fig:TES2}. 
Clearly, the normalized second-order moment can be extracted very accurately, while the normalized moments with orders higher than that cannot be extracted as precisely and the inaccuracy increases with the growing order of $m$.
Moreover, the truncation of the photon number to ten in the measurement resulting from the limited acquisition time
causes a slight but clearly visible systematic error, which diminishes the values gained for $g^{(m)}$. Therefore, in Fig.~\ref{fig:TES2} we restrict ourselves to $g^{(m\le4)}$.
Altogether, our very simple example with Poissonian photon statistics illustrates that with increasing order it becomes more challenging to measure the normalized higher-order moments accurately and that a much more effort is needed to do so.

\section{\label{sec:cv}Accessing the normalized higher-order moments via homodyne detection}

\begin{figure}[t]
\centering \includegraphics[width = 0.4\textwidth]{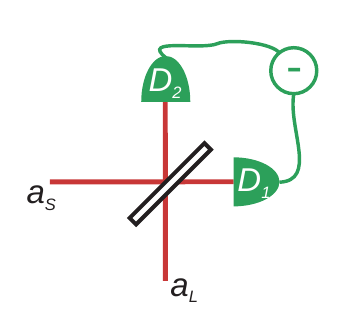}\\
\caption{\label{fig:HDdet} Typical realization of a homodyne detection scheme. Signal and local oscillator fields are interfered on a symmetric beam splitter and detected on a pair of PIN photo diodes $D_1$ and $D_2$. The relative phase between those two fields determines which of the field quadratures will be detected.}
\end{figure}

As an alternative for photon counting, it is also possible to use homodyne detection for accessing the normalized  higher-order moments \cite{Leuchs2017}. In homodyne detection, the signal state of interest with an annihilation operator $\hat{a}_S$ is interfered with a bright local oscillator field having the annihilation operator $\hat{a}_L$ at a symmetric beam splitter and then detected on a pair of PIN photodiodes as shown in Fig.~\ref{fig:HDdet}. The observable of  our choice is the difference current from both diodes, which is proportional to the difference in the detected mean photon number. This is expressed as
\begin{align}
\hat{n}_{-} = \hat{a}^{\dagger}_S\hat{a}_L + \hat{a}^{\dagger}_L\hat{a}_S.
\label{eq:photondiff}
\end{align}
In the limit of a bright local oscillator field, the field operators $\hat{a}_L$ can be replaced by a field amplitude $\alpha_L$ with a relative phase $\theta$ to the signal field. Thus, Eq.~(\ref{eq:photondiff}) can then be written as
\begin{align}
\hat{n}_{-} = \alpha^{\ast}_L \hat{a}_S ~e^{-i\theta}+ \alpha_L \hat{a}_S^{\dagger}~e^{i\theta} = |\alpha_L| \cdot \hat{X}_{\theta},
\end{align}
where $\hat{X}_{\theta} = \hat{a}_S^{\dagger}e^{i\theta} +\hat{a}_S e^{-i\theta}$ is the generalized quadrature operator for arbitrary phases $\theta$.
We note that this definition of the quadrature operator is chosen such that the quadrature variance of the vacuum state, also called the shot noise, equals to unity, which is convient for the homodyne detection. We emphasize that other definitions like the one used in Sec.~\ref{sec:mg}, for which the shot noise takes another value, are possible and, at times, useful.
Additionally, we are only considering a single-mode picture, which is supported by the fact, that the signal field is interfered with the bright local oscillator. This interference realizes an effective modal filter so that the PIN diodes  of the homodyne setup only register the overlapping parts of the signal mode \cite{Raymer1995, Dirmeier2020}. 

Limiting ourselves to the normalized  second-order moment, we substitute the annihilation and creation operators using the expressions $\hat{a} = 1/2~(\hat{X}_{0} + \imath \hat{X}_{\frac{\pi}{2}})$ and $\hat{a}^{\dagger} = 1/2~(\hat{X}_{0} - \imath \hat{X}_{\frac{\pi}{2}})$. Thus, the normalized second-order moment is expressed as \cite{McAlister1997, Grosse2007}
\begin{align}
	g^{(2)} = \frac{\braket{\sum_{\iotaup,\kappaup}{\hat{X}_\iotaup^2\hat{X}_\kappaup^2} - 4\sum_\iotaup{\hat{X}_\iotaup^2} + 12}}{\braket{\sum_\iotaup{\hat{X}_\iotaup^2} -2}^2}
\end{align}
with $\iotaup,\kappaup = \{0, \frac{\pi}{2}\}$. To acquire the necessary quadrature data, one has to either measure the signal state in phase-randomized manner \cite{McAlister1997, Roumpos2013, Lueders2018} or to detect both quadratures, $\hat{X}_0$ and $\hat{X}_{\frac{\pi}{2}}$, simultaneously \cite{Grosse2007, Qi2020}. For the former, one randomly varies the local oscillator phase in Fig.~\ref{fig:HDdet}. To do the latter, it is required to split the input state at a symmetric beam splitter and send the result onto two homodyne detectors with a phase difference of $\frac{\pi}{2}$.
This method can also be extended to the normalized higher-order  moments. Using the same substitution as before, we receive an alternative expression for Eq.~(\ref{eq:gnN})
\begin{align}
g^{(m)} = \frac{\braket{:\hat{N}^m:}}{\braket{:\hat{N}:}^m} = \frac{\braket{(\hat{X}_{0} - \imath \hat{X}_{\frac{\pi}{2}})^m(\hat{X}_{0} + \imath \hat{X}_{\frac{\pi}{2}})^m}}{\braket{(\hat{X}_{0} - \imath \hat{X}_{\frac{\pi}{2}})(\hat{X}_{0} + \imath \hat{X}_{\frac{\pi}{2}})}^m}.
\end{align}

\begin{figure*}[t]
\includegraphics[width = 1.0\textwidth]{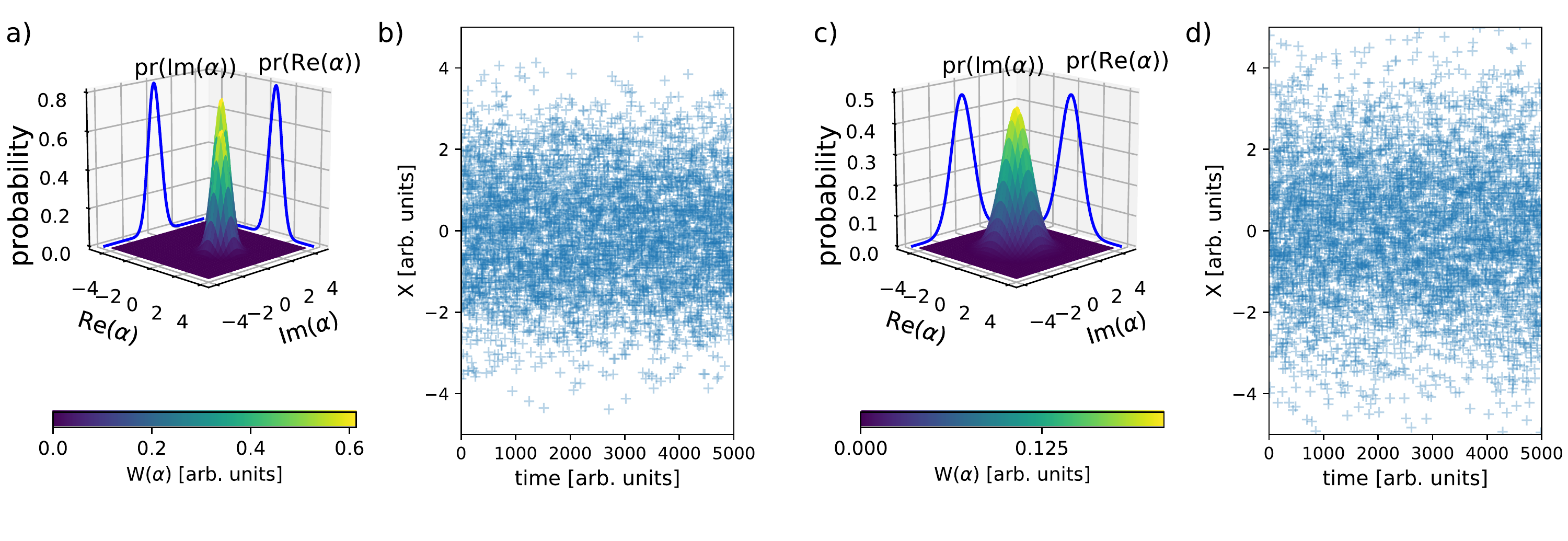}
\caption{\label{fig:WignerHOM} Simulation of the homodyne detection for extracting the normalized higher-order moments.
We illustrate the Wigner functions of (a) a coherent state with its phase aligned along the real axis  and (c) a thermal state that both  have the mean photon number of unity. 
The former is given by $ W(\alpha)  = \frac{2}{\pi} \exp \left (  -2 |1-\alpha|^{2} \right )$ and the latter takes the form $W(\alpha)  = \frac{2}{3\pi} \exp \left (  -2/3 |\alpha|^{2} \right )$ that are both presented here in terms of the complex amplitude $\alpha$ being consistent with the  definition of the Wigner function in Sec.~\ref{sec:mg}. The blue solid lines indicate the probabilities of the marginal distributions, when the Wigner functions are projected on the real $\mathrm{pr}(\textrm{Re}(\alpha))$  and imaginary $\mathrm{pr}(\textrm{Im}(\alpha))$ quadratures.
We emphasize that these quadratures are in analog to $\hat{X}_{0}$ and $\hat{X}_{\frac{\pi}{2}}$ but incorporate another shot noise normalization.
Time traces of the simulated phase-averaged quadrature measurements are presented in (b) for a coherent state like the one in (a) and in (d) for a thermal state like the one in (c) with 5000 samples shown. We notice that in (b) and (d) the shot noise variance is normalized to unity as treated in the main text.}
\end{figure*}

To illustrate the principle, we simulate the phase-randomized detection of coherent and thermal states having a mean photon number of unity and repeat 20 times the simulation with $18\cdot10^6$ samples.
The theoretically calculated Wigner functions and a qualitative depiction of the marginal distributions, which determine the simulated data are depicted in Fig.~\ref{fig:WignerHOM}. The results of calculating the normalized higher-order moments $g^{(m)}$   for our simulated data are shown in table~\ref{tab:g(m)homodyne}. We calculate the different moments by normalizing the measurement data to the shot noise of the local oscillator $\hat{q} = \hat{n}_{-}/\alpha_{L}$ with $\hat{n}_{-}$ being defined in Eq.~(\ref{eq:photondiff}). Assuming a random phase for each measurement, we can calculate the normalized second-order moment as 
\begin{align}
g^{(2)} = \frac{4\cdot \left (\langle\hat{q}^4\rangle - 6 \langle\hat{q}^2\rangle +3\right)}{6\left (\langle\hat{q}^2\rangle-1\right )^2}.
\end{align}
In analogy, the moments of order higher than two can be calculated in a similar manner. The details on deriving these expressions can be found in \cite{Roumpos2013}.

The values for the coherent state stay close to unity as one would expect for a state with a Poissonian photon number distribution. In case of the thermal state, we directly notice the factorial increase in $g^{(m)}$ starting from the value of $g^{(2)} = 2$. We again emphasize that it is important to evaluate the errors, at which the normalized higher-order moments can be extracted. 
The errors increase with growing order of $m$ even in an ideal case of no experimental imperfections taken into account. This is a statistical effect due to the limited size of the measured data set and can be remedied by sampling more data, that is, in an experiment measuring longer. 
However, the increment of the measurement time is only possible if the system under study is stable enough. Often the acquisition time cannot be limitless extended and a trade-off of between statistical errors and other system fluctuations needs to be met.

\begin{table}[!t]
\centering
\caption{\label{tab:g(m)homodyne}Simulated results for the normalized higher-order moments up to the 5th order retrieved 
 from $18\cdot10^6$  samples simulated 20 times for a coherent and thermal state  with mean photon number of unity.}
\begin{tabular}{c|cccc}
\toprule
& $g^{(2)}$& $g^{(3)}$ & $g^{(4)}$& $g^{(5)}$\\
\midrule
coherent state& $1.0000(6)$&$ 0.998(4)$&$1.00(3)$&$1.05(15)$\\
thermal state&$ 1.9991(5)$&$ 5.993(5)$& $23.97(5)$& $120.3(7)$\\
\bottomrule
\end{tabular}
\end{table}

This method has been demonstrated in a variety of optical applications. It was used to monitor the change in coherence properties of the light emitted by a diode laser when passing the lasing threshold \cite{Lueders2018}. Also, in the context of the generation and characterization of squeezed states, it has been used to  probe the higher-order moments of displaced squeezed states \cite{Grosse2007} and to reconstruct the wave function of the state \cite{Han2020}. 
In the particular case of measuring Gaussian states, it is possible to access all characteristics of a quantum state via the quadrature covariance matrix and the mean of the quadrature distributions  as described for example in Ref.~\cite{Olivares2012}. As a consequence, it is also possible to calculate the normalized higher-order moments using the covariance matrix of such Gaussian quantum states \cite{Olivares2018}.
The homodyne detection can also be extended to systems beyond the optical domain and has been demonstrated in the context of characterizing microwave photons in superconducting quantum circuits \cite{Eichler2011}.
\section{\label{sec:states}Classification of quantum states of light}

The normalized second-order moment  is already a standard tool in analyzing the photon-number content of quantum light emitters and is most popularly used for their classification. 
Beyond  that the normalized  moments with orders higher than two  are also becoming essential when gathering information about the nature of the emitted light. Next, we inspect different kind of state classification tasks, in which the normalized higher-order moments have been beneficial.

\subsection{Poissonian and thermal light sources}

\begin{figure}[t]
\centering \includegraphics[width = 0.45\textwidth]{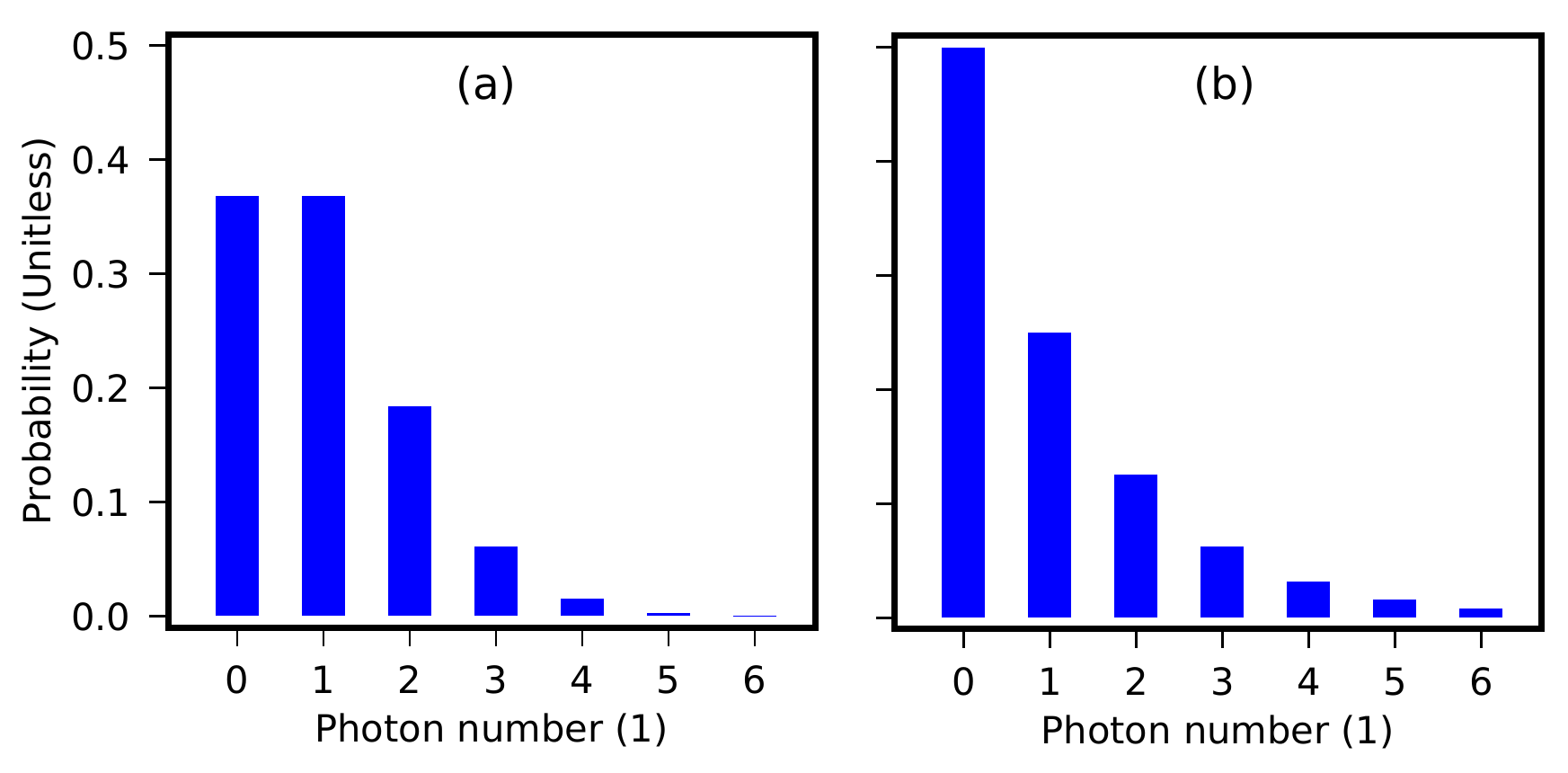}\\
\caption{\label{fig:statcomparison}Photon statistics of (a) Poissonian and (b) thermal states having the mean photon number of unity. The different character of their photon statistics can easily be visually certified.}
\end{figure}

Poissonian and thermal photon statistics play  historically a crucial role and are still today employed in many different quantum optics applications.
The Poissonian photon statistics takes the form
\begin{align}
\rho^{\textrm{Poissonian}}_{n} =  \exp{(-\tilde{n})}\frac{\tilde{n}^{n}}{n!}
\label{eq:poisson}
\end{align}
with $\varg^{(m)}= 1$ as is easily extraced via Eq.~(\ref{eq:gm_stat}), whereas the thermal one can be expressed as
\begin{align}
\rho^{\textrm{thermal}}_{n} =  \frac{\tilde{n}^{n}}{(1+\tilde{n})^{n+1}} = |\xi|^{2n}(1-|\xi|^{2})
\label{eq:thermal}
\end{align}
with $\varg^{(m)}= m!$ delivered by Eq.~(\ref{eq:gm_stat}). In Eq.~(\ref{eq:thermal}) we have presented the two most often used forms for thermal photon statistics connected via $|\xi|^{2} =\frac{\tilde{n}}{1+\tilde{n}} $, which is usually called the strength of the thermal state.
In Eqs (\ref{eq:poisson}) and (\ref{eq:thermal}) $\tilde{n}$ denotes the mean photon number of the state. In Fig.~\ref{fig:statcomparison}(a--b) we illustrate the Poissonian and thermal photon statistics having the mean photon number of unity. Clearly, there are distinct differences between the statistics that are also reflected in their respective higher-order moments. The thermal statistics always has a non-zero vacuum contribution, which takes the highest value in the distribution, whereas that of the Poissonian one gradually vanishes when increasing mean photon number. Thermal photon statistics thus is wider, which reflected by the variance, and has a stronger tilt (skewness) and longer tail (kurtosis) that can mathematically be associated with the second-, third- and fourth-order moments, respectively \cite{Mandel1996}. Therefore, the higher-order moments can  directly reveal important figures of merit of the shape of the studied photon-number distribution.

The measurements of the normalized moments have become a powerful tool in understanding the origin of the lasing in nano- and microstructured devices by revealing the difference between thermal statistics below the laser's threshold and Poissonian one above it \cite{Assmann2009}.
Through such investigations one has gained access to the importance of different parameters in lasing structures, such as the light confinement in a cavity with a high quality factor or the enhancement experienced via small mode-volumes \cite{Strauf2006, Ulrich2007}. In thresholdless lasers that have a high rate of spontaneous emission into the lasing mode,  the pronounced kink in the laser's input-output curve disappears \cite{Noda2006}. In that case one can classify the emitted radiation via its photon statistics, while also observing the effect of the cavity parameters on the behavior of the conventional input-output characteristics \cite{Chow2018, Reeves2018}.

More insight of the emitted light can be gained by measuring also the normalized moments with orders higher than two, for example, in order to extract the degree, to which the investigated light resembles that from a chaotic source i.e.,~thermal source or from a laser source \cite{Elvira2011, Rundquist2014, Schlottmann2017}. 
The measurements of normalized  higher-order moments have also become expedient for exploring the internal dynamics of nanolasers \cite{Wiersig2009} or more sophisticated photon-number effects like superradiance accompanied with photon bunching that is stronger than for the thermal light \cite{Jahnke2016, Wang2020}.
Altogether, these moments are expedient in resolving the quality of the light emitted from the miniaturized laser sources \cite{Dynes2011}.

Thermal light plays a crucial role both in the optical  \cite{Zhang2005, Boitier2009, Paleari2004} and microwave regimes \cite{Goetz2017}.
Pseudo-thermal light is usually produced by sending laser light through a rotating ground disk, after which a single speckle is collected via a pinhole \cite{Arecchi1965}. The action of this experimental arrangement can be generalized to other light sources also \cite{Li2020}. Interestingly, the properties of photon-number subtracted thermal states show a gradual transition towards Poissonian statistics indicating a drop in the photon-number correlations \cite{Y.Zhai2013, Katamadze2020}. 
Undoubtedly, both true- and pseudo-thermal light sources have also become important in the large field of quantum imaging  that is based on the measurements of photon correlations \cite{Zhang2005, Valencia2005}.
In fact, the utilization of the higher-order correlations of thermal light \cite{Chan2009} can result in the increased  quality of the ghost images \cite{Chen2010, Chan2010, Iskhakov2011}. 
Overall, the dissimilar behavior of the normalized higher-order moments in the different classes of light allow one even to recognize and resolve the mode occupation in complex optical states \cite{Goldschmidt2013}. Moreover, the higher-order moments of other exotic light sources such as the free-electron laser  emitting at the extreme ultraviolet regime have been measured to understand its spatio-temporal properties and photon-number characteristics in terms of spectral filtering \cite{Singer2013}.

\subsection{Characteristics of single-photon emitters}

A perfect $k$-photon emitter incorporates only the $k$-th photon contribution and has the photon statistics 
\begin{align}
\rho^{\ket{k}}_{n} = \left \{ \begin{array}{cc} 1 & \textrm{if \quad} n= k  \\ 0 & \textrm{otherwise}\end{array} \right .,
\label{eq:one_photon}
\end{align}
where $k$ is belonging to the natural numbers.  Using Eq.~(\ref{eq:gm_stat}) the $m$-th order factorial moment of the photon statistic in Eq.~(\ref{eq:one_photon}) takes the form
\begin{align}
\varg^{(m)} = \frac{m(m-1)\dots(m-k+1)}{k^{m}}.
\label{eq:photon_gm}
\end{align}

Thus, for the \emph{ideal  single-photon emitter} that has $k = 1$ all the values $\varg^{(m\ge2)}= 0$. Therefore, the vanishing normalized second-order moment is well adapted for evidencing that all photon-number contributions higher than or equal to two disappear. Due to the experimental imperfections it is very challenging to reach this value, and  therefore, often a value $\varg^{(2)}=  0.5$, which is valid for an ideal two-photon state [$\varg^{(2)} = 1-\frac{1}{k}$  with $k=2$ in Eq.~(\ref{eq:photon_gm})], is regarded as the upper limit for the single-photon emitters \cite{Chavez-Mackay2020}. 

\begin{figure}[t]
\centering \includegraphics[width = 0.45\textwidth]{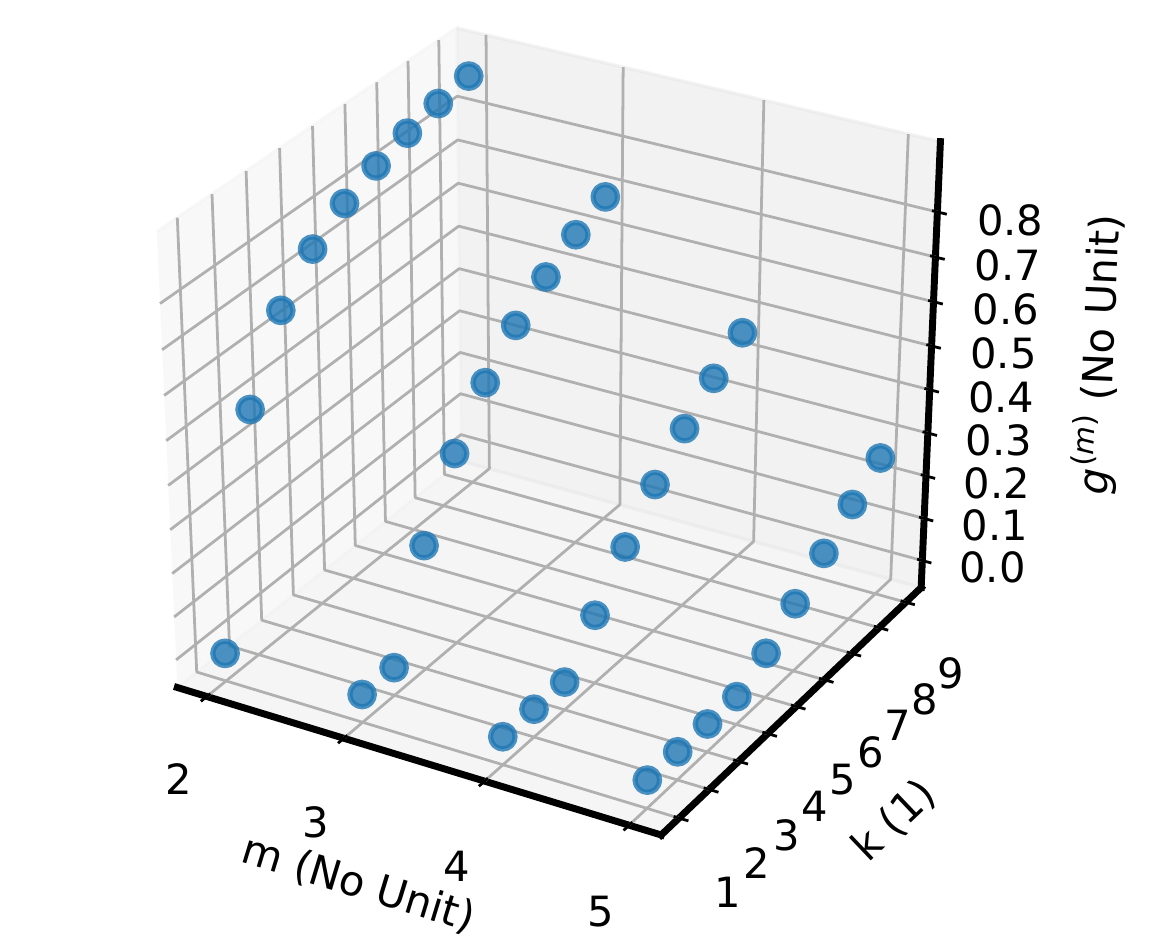}\\
\caption{\label{fig:gmk}The behavior of the normalized moments $\varg^{(m)}$ up to the 5th-order ($m = 2\dots 5$) for photon number states in terms of the photon number $k$.}
\end{figure}

Nevertheless, all values $g^{(2)} <1$ hint at a sub-Poissonian light source and verify its non-classicality. However, we emphasize that the normalized second-order moment cannot distinguish the vacuum contribution of the light source. Therefore, the extraction efficiency or the state fidelity with respect to the one-photon state is often reported alongside \cite{Gruenwald2019}. This way, the single photon characteristics have been demonstrated from the emission from atoms and ions \cite{Kimble1977, Grangier1986, Diedrich1987} and from the solid state including at least quantum dots \cite{Michler2000, Dauler2008, Arakawa2020}, silicon- and nitrogen-vacancy centers in diamond \cite{Jelezko2006, Neu2011, Steudle2012}, cold and room-temperature molecules \cite{Lounis2000, Toninelli2020} and microwave circuits that can be regarded as artificial atoms \cite{Bozyigit2011,Gu2017}. Further heralding of single photons has been demonstrated via the non-linear optical processes of four-wave mixing \cite{McMillan2013, Faruque2019} and PDC [see Sec.~\ref{sec:bistates}].

Due to the fact that true single photons are difficult to prepare in the laboratory and that the prepared states most often incorporate experimental imperfections, such as higher-photon-number contributions, it is also worthwhile to examine their normalized moments with orders higher than two.
Such measurements have been employed in several experiments in order to more deeply understand their characteristics.
In order to  estimate the ratio of the single-photon emission to the parasitic background light such investigations have been conducted at least up to the third order with quantum dots \cite{Stevens2014} and up to the second order with ensembles of emitters based on nitrogen-vacancy centers \cite{Moreva2017}. Further, the normalized moments 
of the emission from colloidal two-dimensional nanocrystals  have  been investigated upto the fourth order  in order to analyze their behavior in terms of the size and emission wavelength of the emitters \cite{Amgar2019}. Additionally, measurements of normalized moments up to the fourth order with light emitted from ensembles of colloidal quantum dots acting as single-photon emitters have been investigated to explain the effect of the cluster size on the emitted light characteristics and its non-classicality  \cite{Qi2018}.

\subsection{Photon-pair states and other multi-photon states}

Light sources producing photons in pairs incorporate strong photon-number correlation between the created photons. 
The most famous state including only even photon-number contributions probably is the squeezed vacuum, usually created by non-linear optics via either PDC or four-wave mixing \cite{Slusher1985, Wu1986}. The strong correlation between the created photon pairs can be verified by measuring the normalized second-order moment \cite{Ivanova2006, Kuga1993, Boitier2011}. Displacing the squeezed vacuum in phase-space produces quadrature squeezed states of light and results in interesting features in the state's characteristics like for example on sub-Poissonian behaviour and clear non-classical features that have been probed by measuring the second-order normalized moment \cite{Grosse2007,Han2020,Olivares2018,Koashi1993}. 
However, twin-photon states can also be created from the emission from solid state. Bunching higher than that in thermal light has been reported by measuring the normalized second-order moment from a twin-photon cascade emitted from quantum dots \cite{Heindel2017}.
Also correlations between photon triplets emitted from a solid state source have been verified \cite{Khoshnegar2017}.  
Moreover, quadrature squeezed light can also be prepared in quantum-dot systems \cite{Schulte2015}.

The generation of more sophisticated states like path-entangled NOON-states, describing  a non-separable superposition  $\propto \ket{\mathcal{N} 0}+\ket{0\mathcal{N}}$ with $\mathcal{N}$ photons distributed between two independent paths, often incorporates preparation of photon-pair states and  their measurements with correlation functions  \cite{Mitchell2004, Afek2010}.  
Also other exotic states such as Schr\"odinger Kitten states defined as a superposition of two coherent states with opposite phases
can be created with photon subtraction from squeezed vacuum states \cite{Neergaard-Nielsen2006, Ourjoumtsev2007,Wakui2007}.
The generation of other superposition states, such as that of the vacuum and single-photon displaced in the phase space also often relies on the squeezed state generation and photon-number subtraction \cite{Zavatta2004, Coelho2016}.

\subsection{\label{sec:bistates}Characteristics of bimodal states of light}

Especially interesting are bimodal states of light, in which photon correlations are generated between two distinct modes of light.
The characteristics of such states can be measured for example by accessing the joint photon-number distribution of the state with two true photon-number resolving detectors. 
Detection with two TESs has been employed investigating the characteristics of such states created at least  in bimodal microlasers that incorporate strong anti-correlation between the created modes of light \cite{Schmidt2021}
and  via twin-beam production from PDC emission, which in contrast results in strong photon-number correlation between the created modes \cite{Harder2016, Sperling2017, Magana-Loaiza2019, Laiho2019, Tiedau2019}.

\begin{figure}[!b]
\centering \includegraphics[width = 0.45\textwidth]{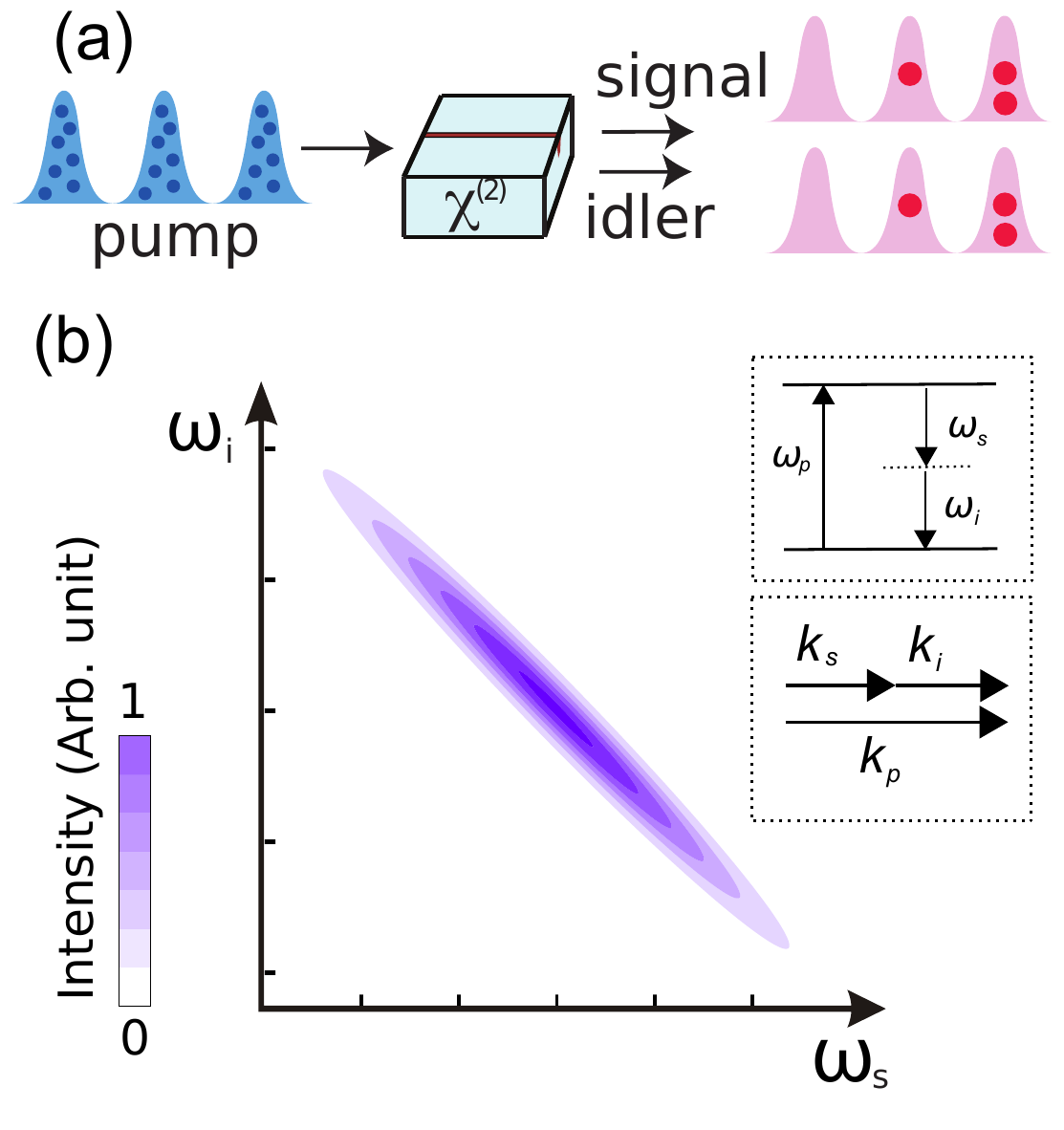}
\caption{\label{fig:PDC} Illustration of typical PDC characteristics in a collinear configuration. In (a) a laser beam is coupled in the non-linear optical material with a second-order non-linearity $\chi^{(2)}$, through which the pump ($p$) photons split in signal ($s$) and idler ($i$) photons propagating parallel to each other. Such a scheme can be realized both in waveguided and bulk configurations. If signal and idler are created in separable modes in the spectral, spatial or polarization degree of freedom, twin beams are generated. In (b) a Gaussian approximation of  typical joint spectral properties, that is $|f(\omega_{s}, \omega_{i})|$ from Eq.~(\ref{eq:JSA}), is demonstrated, which incorporates a strong anti-correlation in the frequency space. The exact shape of the joint spectral correlations of signal and idler ordinarily stems from the dispersion properties of the used material \cite{W.P.Grice1997, Zielnicki2018, Gianani2020}. Such strong spectral correlations result in multimodal PDC emission. The insets show the conservation laws for energy and momentum, the latter of which is described in terms of the wavenumber $k_{\zeta} = \frac{\omega_{\zeta}}{c}n_{\zeta}$ ($\zeta= s, i, p$) with $c$ being the speed of light and $n_{\zeta}$ the refractive index of the corresponding mode. Natural phasematching is achieved if $k_{p} = k_{s}-k_{i}$, which is possible for example in the semiconductor Bragg-reflection waveguides basing on modal phasematching \cite{Laiho2016}.
In periodically poled waveguides and crystals a wavevector mismatch ($k_{p}-k_{s}-k_{i} \ne 0$) can be compensated with an artificial wavevector component that can be used for example for frequency tuning \cite{Fiorentino2007, Eigner2020}.}
\end{figure}

In the following, we focus on the properties of  twin beams generated from PDC emission as illustrated in Fig.~\ref{fig:PDC}
and concentrate on the conventional case, in which the PDC process is pumped with a coherent laser source. We investigate the properties of the joint higher-order normalized moments of the signal and idler beams that in general  provide several means for accessing the intrinsic PDC process parameters. 
Recently, it was shown that the characteristics of the twin beams change if the pump light has other than Poissonian characteristics \cite{Meher2020}. However, until today only very few experiments exploit other kind of pump light than a laser like for example heralded single photons from PDC \cite{Huebel2010} or four-wave-mixing \cite{Ding2015}.

The joint higher-order correlations can be investigated in analog to Eq.~(\ref{eq:q_N}) as \cite{Christ2011}
\begin{align}
& 
{g}^{(w,\upsilon)} = \frac{\braket{:(\sum_{q}\hat{A}^{\dagger}_{q}\hat{A}_{q})^{w}(\sum_{q^{\prime}}\hat{B}^{\dagger}_{q^{\prime}}\hat{B}_{q^{\prime}})^{\upsilon}:}}{\braket{\sum_{q}\hat{A}^{\dagger}_{q} \hat{A}_{q}}^{w} \braket{\sum_{q^{\prime}}\hat{B}^{\dagger}_{q^{\prime}}\hat{B}
_{q^{\prime}}}^{\upsilon} },
\label{eq:GlauberG}
\end{align} 
in which $A^{\dagger}_{q}$ ($A_{q}$)  and $B^{\dagger}_{q^{\prime}}$ ($B_{q^{\prime}}$) denote the broadband photon creation (annihilation) operators in signal ($s$) and idler ($i$) modes labelled with $q$ and $q^{\prime}$, respectively, and the indices $w$ and $\upsilon$ label the order of the joint moment.

The two-mode squeezed state $\ket{\Phi}_{s,i}$  of signal and idler, can be expressed in terms of the unitary squeezing operator $\hat{S}_{s,i}$ as \cite{Christ2011}
\begin{align}
\ket{\Phi}_{s,i} =  \hat{S}_{s,i}\ket{0}  = \textrm{e}^{\sum_{q} r_{q} \hat{A}_{q}\hat{B}_{q} -h.c.}\ket{0} ,
\label{eq:state_tb}
\end{align}
in which the parameter $r_{q} = \mathcal{B}\lambda_{q}$ is defined in terms of the process strength  $\mathcal{B}$ and the real-valued weight of the $q$-th mode $\lambda_{q}$ and $h.c.$ denotes the hermitian conjugate. These weights obey the normalization $\sum_{q}\lambda^{2}_{q}= 1$. The effective mode number can be extracted via $\mathcal{K} =1/ \sum_{q}\lambda^{4}_{q}$ and the value $\mathcal{K} = 1$ corresponds to the single-mode (SM) PDC emission, while $\mathcal{K} \gg 1$ denotes multi-mode (MM) PDC emission \cite{Eberly2006}. 
To this end, the first SM PDC source was experimentally demonstrated in Ref.~\cite{P.J.Mosley2008}.

The joint spectral properties of signal and idler are determined by their joint spectral amplitude, which is connected to these weights via
\begin{align}
f(\omega_{s}, \omega_{i}) = \sum_{q}\lambda_{q} \ \varphi_{q}(\omega_{s}) \ \phi_{q}(\omega_{i}).
\label{eq:JSA}
\end{align}
The frequency-dependent sets of the basis functions $\{\varphi_{q}(\omega_{s})\} $ and $\{\phi_{q}(\omega_{i})\}$ for signal and idler, respectively, both obey the conditions in Eq.~(\ref{eq:basis})  and can be accessed via the Schmidt-mode decomposition accomplished as the singular-value decomposition.

\begin{figure}[!tb]
\centering \includegraphics[width = 0.45\textwidth]{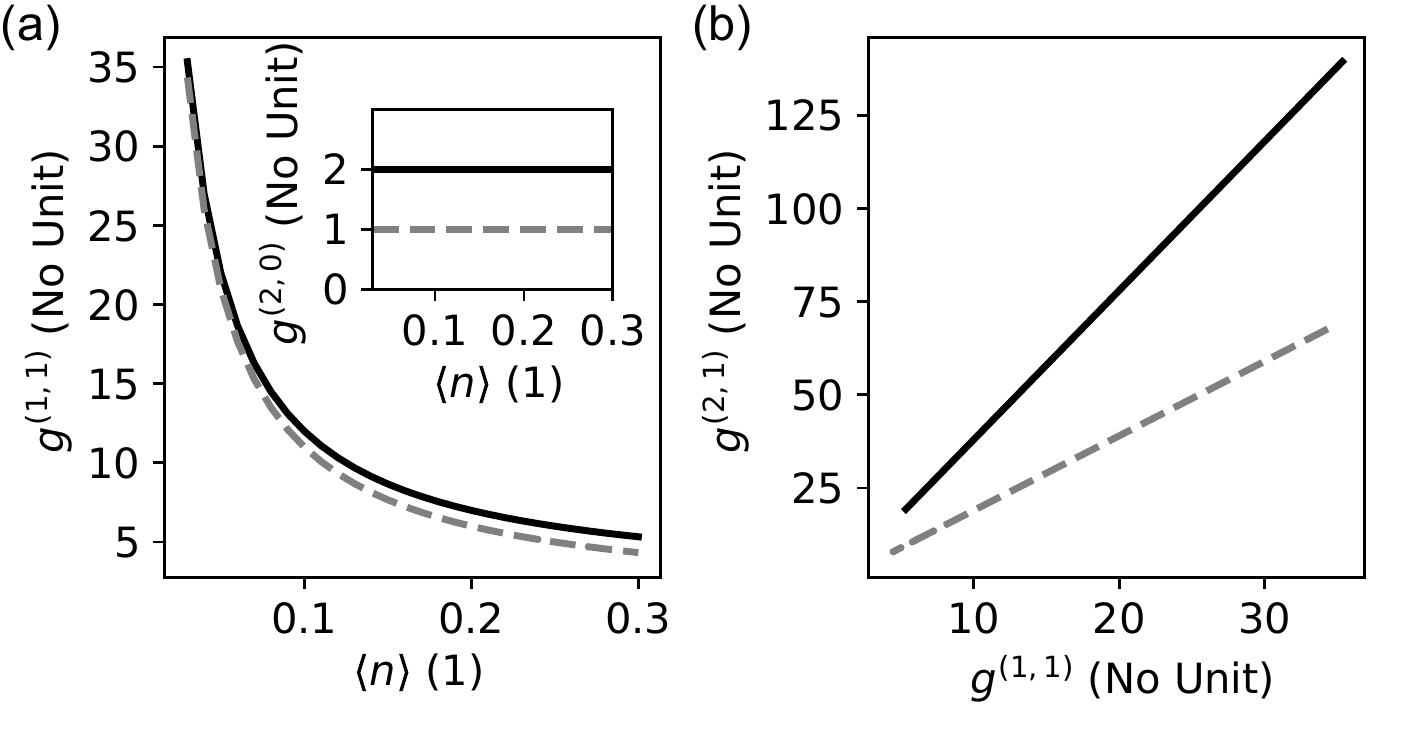}
\caption{\label{fig:gmPDC}  Characteristics of the joint normalized moments of twin beams with the lowest orders. In (a) we illustrate the inverse proportionality between the CAR represented by $g^{(1,1)} $ and the PDC mean photon number. In (b) we present the linear dependence between $g^{(2,1)}$ and $g^{(1,1)}$.   The solid black lines represent the characteristics of SM PDC process, whereas the properties of highly MM PDC calucalated with $\mathcal{K}\rightarrow \infty $ are illustrated with gray dashed lines. The inset in (a) shows the behavior of $g^{(2,0)}$ with respect to the PDC mean photon number.}
\end{figure}

In order to evaluate the joint normalized moments of signal and idler, we employ the transformations  \cite{S.M.Barnett1997}
\begin{align}
\label{eq:SAS}
 &\hat{S}^{\dagger}_{s,i} \hat{A}_{q}  \hat{S}_{s,i}  =\textrm{cosh}(r_{q})\hat{A}_{q} +  \textrm{sinh}(r_{q}) \hat{B}_{q} \quad {\textrm{and}} \\
 &\hat{S}^{\dagger}_{s,i} \hat{B}_{q}  \hat{S}_{s,i}  = \textrm{cosh}(r_{q})\hat{B}_{q} + \textrm{sinh}(r_{q})\hat{A}_{q},
 \label{eq:SBS}
\end{align}
that describe the optical squeezing process. The mean photon number of the PDC emission can easily be evaluated via $\braket{n} = \sum_{q^{\prime}} \braket{\hat{B}^{\dagger}_{q^{\prime}}\hat{B}_{q^{\prime}}}  = \sum_{q} \braket{\hat{A}^{\dagger}_{q}\hat{A}_{q}} = \sum_{q} \textrm{sinh}^{2}(r_{q}) \approx  \mathcal{B}^{2}$. The approximation in the last step is valid for MM PDC in the limit of weak squeezing, for which  $\textrm{sinh}(r_{q}) \approx \mathcal{B}\lambda_{q}$ applies.
Additionally, in that region, that is, at low pump powers, the evaluation of the lowest orders of the joint normalized moments delivers \cite{Christ2011}
\begin{equation}
{g}^{(1,1)} \approx\frac{1}{\mathcal{B}^{2}}+\sum_{q} \lambda^{4}_{q} + 1 = \frac{1}{\braket{n}}+\frac{1}{\mathcal{K}}+1,
\label{eq:CAR}
\end{equation}
\begin{equation}
{g}^{(2,0)} = 1+\sum_{q} \lambda^{4}_{q}  = 1+\frac{1}{\mathcal{K}} \quad \textrm{and}
\label{eq:g20}
\end{equation}
\begin{equation}
{g}^{(2,1)} = \bigg [ 1+ \frac{2}{\braket{n}} \bigg] {g}^{(2,0)}+ 2 \bigg [ \frac{1}{\mathcal{K}} + \sum_{q}\lambda_{q}^{6}\bigg].
\label{eq:g21}
\end{equation}
All the joint normalized moments provide useful insight in to the characteristics of the signal and idler beams. The most often used joint normalized moment is probably $g^{(1,1)}$ in Eq.~(\ref{eq:CAR}), which corresponds to the well-known coincidences-to-accidentals ratio (CAR) \cite{Friberg1985, Kono1996}. It can directly and loss-independently be employed for estimating the mean photon number of the PDC state as shown in Fig.~\ref{fig:gmPDC}(a) after the effective number of modes is extracted for example via Eq.~(\ref{eq:g20}). 
The inverse proportionality between $g^{(1,1)}$ and the mean photon number is evident \cite{Tanzilli2002, Zhu2013}. In case of MM PDC the estimation of the mean photon number from CAR is straightforward  \cite{Guenthner2015}. High values of CAR are desired in order to prove a strong photon-pair correlation between signal and idler, such that spurious counts are successfully  suppressed \cite{Bocquillon2009, Krapick2013, He2015, Kultavewuti2016, Chen2018}. 

The normalized second-order moment of an individual twin beam, that is, $g^{(2,0)}$ in Eq.~(\ref{eq:g20}), provides a direct access to the \emph{effective number} of the spectral modes excited. 
Most interestingly, the value of $g^{(2,0)}$ for a marginal beam i.e., signal or idler, is in general independent of the pump power, especially in waveguided PDC. 
When investigating degenerate PDC, that is, signal and idler lie at the same wavelength, it is also utterly important to experimentally verify the independence of $g^{(2,0)}$ from the pump power  \cite{Laiho2011}. Namely, even a slight leakage of the signal beam to the idler beam path or vice versa produces a background of single-beam squeezing, which causes strongly pump power dependent values \cite{Christ2011, Wakui2014} and can result in a wrong interpretation of the effective mode number.
A more generalized analysis of the PDC process has shown that the photon-number properties of the marginal beams can be manipulated by the photon-number properties of the pump. In general, $g^{(2,0)}$  can be twice as high than the normalized second-order moment of the pump \cite{Meher2020}.

The number of the excited modes in the PDC emission plays a crucial role when deciding, to which quantum optics tasks the PDC source in question is suitable for. 
Ideally, if the PDC is emitted into a single spectral mode, which is suitable for heralding photon-number Fock states from the PDC emission, the photon-number properties of the marginal beams show thermal behavior \cite{Paleari2004, Vasilyev2000}. In the case of a MM PDC  process, which is suitable for example for entanglement generation or spectrally multiplexed states \cite{Alibart2016, Kues2017}, the photon-number properties of the marginal beams follow that of Poissonian light \cite{Haderka2005, Avenhaus2008}. Several different PDC process parameters can be used to tune the effective mode number such as the pump bandwidth \cite{Eckstein2011, Harder2013, Bruno2014}. Also the pump depletion has been shown to affect the spectral mode profiles and thus  dramatically modify the multimode structure of twin beams \cite{Allevi2014}. In particular systems, the number of exited modes can be engineered and a direct selection of the desired spatial or temporal modes has become possible \cite{Perez2014, Foertsch2015, Ansari2018, Graffitti2020}. 
Apart from that spectral filtering provides simple means for reducing the mode number of the twin beams \cite{Tapster1998}. 

The joint normalized moments with other orders such as $g^{(1,2)}$ in Eq.~(\ref{eq:g21}) also play a crucial role in understanding the intrinsic behavior of the PDC emission. The measurement of the joint normalized moments $g^{(w,\upsilon)}$ is possible directly via the measured joint photon-number distribution of signal and idler \cite{Harder2016, Magana-Loaiza2019, Laiho2019} or via manifold coincidence counting after sending both signal and idler beams in a spatially or temporally resolving beam splitter networks \cite{Avenhaus2010, Laiho2011}. In the latter case, the joint normalized moments are evaluated by dividing the probability of a coincidence click between $w$ selected detectors in signal arm and $\upsilon$ ones in idler arm with a product of the probabilities of detecting a single click in each of these selected detectors. 
The joint normalized photon correlations of the PDC emission grow with the increasing orders of $w$ and $\upsilon$ as illustrated in Fig.~\ref{fig:gmPDC}(b) \cite{Avenhaus2010, Harder2016, Laiho2011, Allevi2012}.  The normalized moments of higher orders better reveal the interplay of  the effective modes in the PDC emission \cite{Laiho2011, Wakui2014}, and apart from providing information of the strength of the correlations between the higher photon numbers, they have been used to verify the non-classicality of the twin beams \cite{Avenhaus2010, Allevi2012}.

\begin{figure}[!tb]
\centering
\includegraphics[width = 0.4\textwidth]{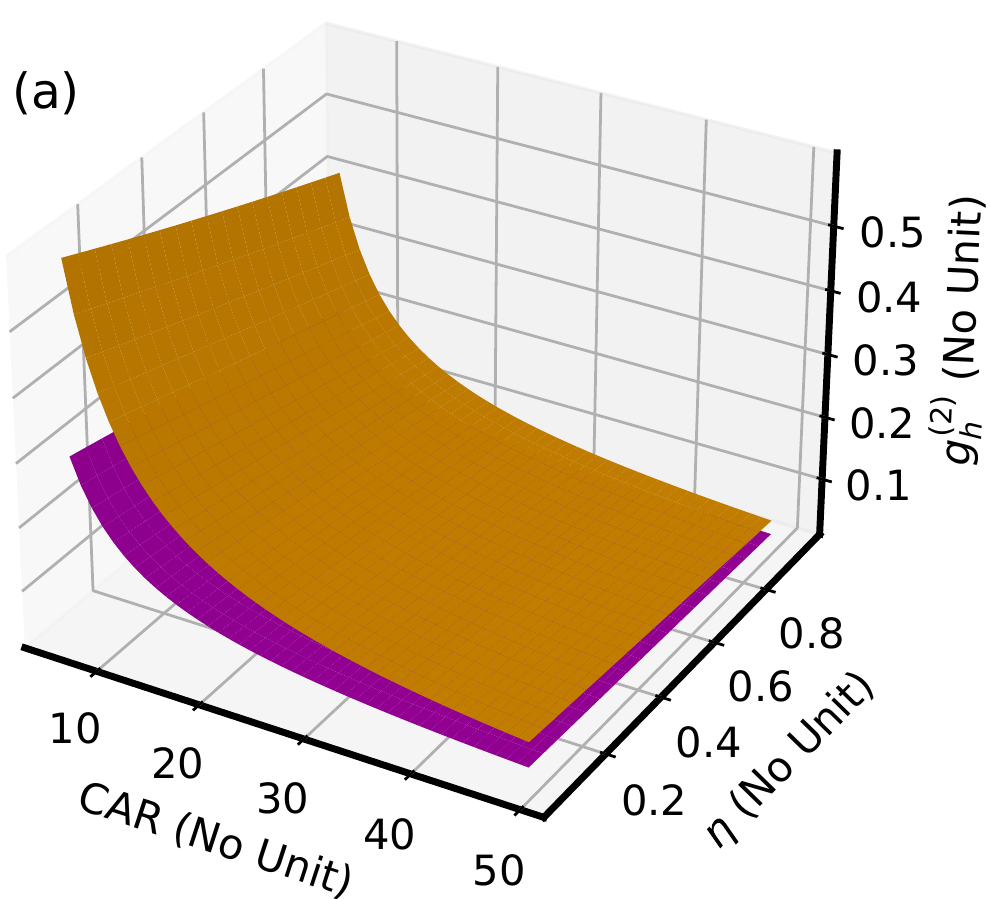}
\includegraphics[width = 0.4\textwidth]{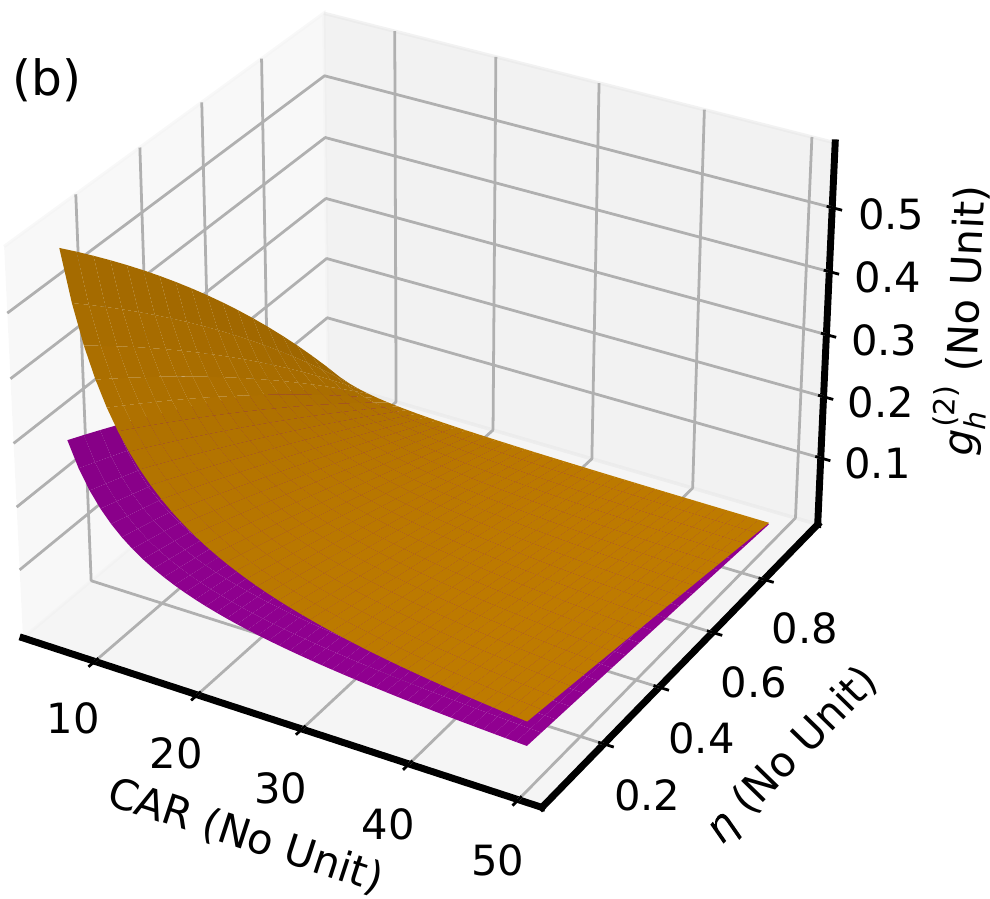}
\caption{\label{fig:g2h} The properties of ${g}^{(2)}_{h}$ in terms of  CAR  and the efficiency of the heralding detector $\eta$ for both SM (orange) and MM (magenta) PDC when (a) a click detector and (b) true photon-number-resolving detector  is used for heralding. 
For the calculation we start with the joint signal-idler state $\ket{\Psi} = \sum_{n}\Lambda_{n}\ket{n,n}_{si}$, in which $|\Lambda_{n}|^{2}$ describes the photon statistics of the strictly photon-number correlated signal and idler and corresponds to thermal statistics in SM PDC and Poissoinian one in MM PDC. The density matrix of the heralded state is estimated via $\varrho_{s} = \frac{1}{\Upsilon} \textrm{Tr}_{i}\left \{(\hat{1}_{s} \otimes \hat{\Pi}^{\mathcal{O}}_{i}) \ket{\Psi}\bra{\Psi} \right \}$, in which $\Upsilon$ accounts for the normalization of the heralded state and $\hat{\Pi}_{i}^{\mathcal{O}}$ denotes the detection operation at the heralding with $\mathcal{O}$ labeling the detector type \cite{ph_det}. The second-order normalized moments are thus evaluated from the heralded state photon statistics given by $\rho_{n} = \textrm{Tr}_{s}\left \{ \varrho_{s}\ket{n}_{s, s}\hspace{-0.5ex}\bra{n} \right \}$ via Eq.~(\ref{eq:gm_stat}).}
\end{figure}

Altogether, the twin beams are ideal for state manipulation and often employed for example in the preparation of heralded number states \cite{A.I.Lvovsky2001, A.Ourjoumtsev2006}. 
However, the quality of the heralded state is strongly affected e.g.~by the losses in the herald and the photon-number resolution of the heralding detector  \cite{Branczyk2010, Laiho2019,  Tiedau2019, DAuria2012, Rohde2015}. 
Therefore, most often the target states are restricted to the heralded single photons.
The normalized second-order moment  conditioned on the detection of the herald, often labelled with ${g}^{(2)}_{h}$ suits well for the classification of the these states. 
Additionally, the heralded state characteristics is connected to the underlying joint photon-number distribution of signal and idler, and the values reached for ${g}^{(2)}_{h}$ depend on the higher photon-number contributions of the joint signal-idler state. Following Ref.~\cite{Laiho2019} we illustrate in Fig.~\ref{fig:g2h} the achievable values for ${g}^{(2)}_{h}$ in terms of CAR aka $g^{(1,1)}$ of the PDC emission and Klyshko efficiency \cite{Klyshko1996} of the herald for both SM and MM PDC when detecting the herald with either a click detector or a true photon-number-resolving detector. One benefits of the heralding with true photon-number-resolving detection only at high detection efficiencies. Moreover, the intrinsic PDC process parameters strongly affect the residual higher photon-number contributions in the heralded state \cite{Laiho2011}. 
Apart from the platforms on bulk crystals \cite{Hong1986}, recently several integrated optics devices have shown close-to-zero values for ${g}^{(2)}_{h}$ ranging from ferro-electric platforms \cite{URen2005, Ngah2015} to semiconductor nanowires and waveguides \cite{Guo2017, Belhassen2018}. Heralding of higher photon-number states similarly greatly suffers from the losses experienced in the heralding arm \cite{Waks2006, Tiedau2019, Sperling2015, Hlousek2019}. For those states, one no longer can just decrease the pump power in order to get to higher CAR values since the yield then dramatically drops.

\section{\label{sec:noncl}Nonclassicality measures and higher-order moments}

The normalized higher-order moments can reveal information of the quantum state's characteristics in phase-space even without having the complete access to its  density matrix  or to its alternative representations, the quasi-distributions functions. 
One of these quasi-distribution functions, the $P$-representation, in other words the normally-ordered expansion of the density matrix, is given for a single-mode state by
\begin{align}
\varrho(\alpha) = \int d\alpha^{2}P(\alpha)\ket{\alpha}\bra{\alpha}
\end{align}
and defined in terms of coherent states having the amplitude $\alpha$.
For a classical state the $P$-representation is well-behaving and takes only non-negative values, in other words, it can be interpreted as a classical probability distribution. Singularities in this function  or negative values are signs of \emph{non-classicality}.

Several non-classicality criteria based on the $P$-representation aim at revealing cases, in which it cannot anymore be interpreted as a classical probability distribution. Luckily, these criteria can often be formulated and observed via the higher-order moments through the definition  \cite{Lee1990, Klauder2006}
\begin{align}
\braket{\hat{\textbf{a}}^{\dagger m}\hat{\textbf{a}}^{m^{\prime}}} = \int d\alpha^{2}P(\alpha)\alpha^{*m}\alpha^{m^{\prime}}.
\label{eq:asym_gm}
\end{align}
In a compact way, one can formulate the non-classicalities with matrices of these moments \cite{Agarwal1992,Miranowicz2010, Sperling2013}. In their simplest form such a matrix can be defined as
\begin{align}
\mathcal{M} = \left [
 \begin{array}{ccccc} 
g^{(0)} & g^{(1)} & g^{(2)}  & \cdots & g^{(m-1)} \\ 
g^{(1)} & g^{(2)} & g^{(3)} & \cdots & g^{(m)} \\ 
g^{(2)} & g^{(3)} & g^{(4)} & \cdots & g^{(m+1)} \\ 
\vdots &\vdots &\vdots &\ddots&\vdots \\
g^{(m-1)} & g^{(m)} & g^{(m+1)} & \cdots & g^{(2m-2)} \\ 
\end{array} \right ].
\label{eq:detmxm}
\end{align}
For a classical state, whose $P$-representation does not take any negative values, all the eigenvalues of the matrix in Eq.~(\ref{eq:detmxm}) are positive. Thus, searching for negative values within the eigenvalues or finding a negative determinant for it is a sign of the non-classicality. Moreover, for Poissonian photon statistics the determinant of such a matrix takes the value of zero, which can be used as a reference value. Several experiments have verified in such manner the non-classicality of the joint signal-idler states \cite{Magana-Loaiza2019, Sperling2015, Arkhipov2016} as well as  that of the heralded photon-number states \cite{Sperling2017} and clusters of single-photon emitters \cite{Bohmann2019}.

Already, the well-known condition $g^{(2)}<1$, is one of such non-classicality criteria directly probing the non-classicality of the $P$-representation of a single-mode state \cite{Titulaer1965, Klauder2006, Agarwal1992}. However, quantum states may show a transition between sub- and super-Poissonian photon statistics. Such effects have been measured in the phase-space at least for the displaced squeezed states \cite{Grosse2007}, displaced single-photon states \cite{Laiho2012} and photon-catalyzed optical coherent states \cite{Bartley2012}. 
Moreover, when looking more closely  for example on the single photons displaced in phase space, at higher displacements their $g^{(2)} > 1$ and the normalized second-order  moment is no longer adequate for proving their non-classicality \cite{Laiho2012}. Indeed, in some cases a more general form of this non-classicality criteria given by 
\begin{align}
g^{(m+1)}< g^{(m)}
\label{eq:qm_ineq}
\end{align} 
\cite{Titulaer1965, Klauder2006} can be useful for proving the non-classicality.  

Another interesting non-classicality criterion is based on Schwarz inequality for the $P$-representation, and it can be expressed in the form \cite{Titulaer1965, Klyshko1996a}
\begin{align}
{g}^{(h-m)}{g}^{(h+m)} \ge \left [{g}^{(h)} \right]^{2}
\label{eq:Schwarz}
\end{align}
with $h$ and $m$ used for denoting the required orders for the moments.
This criterion has also been investigated both with heralded single photons \cite{Waks2006a, Allevi2012} and single photons from solid state emitters \cite{Qi2018} to find non-classical features in spite of experimental imperfections in the studied states.

Many other non-classicality criteria basing on the $P$-representation for multi-mode states such as the twin beams have also been derived in terms of the joint normalized higher-order moments $g^{(w,\upsilon)}$ \cite{Miranowicz2010, Vogel2008}.  The PDC emission producing twin beams have been experimentally shown to obey such conditions \cite{Avenhaus2010}. 
One of their simplest form is $g^{(1,1)} > \sqrt{g^{(2,0)}\ g^{(0,2)}}$  resulting in the condition $\textrm{CAR}>2$ for the twin beams generated from SM PDC, which is  pumped with well-behaving laser \cite{Loudon2000, Miranowicz2010, Vogel2008,  Reid1986}.
Many quantum optical states exhibit also asymmetric moments as described in Eq.~(\ref{eq:asym_gm}) with $m \neq m^{\prime}$ and providing information of the phase characteristics of the investigated state. Until, today such asymmetric moments have been extracted only with homodyne tomography \cite{Eichler2011}. Nevertheless, it is possible to access them directly by measuring the higher-order moments of states displaced in phase space by varying only the phase of the displacement, while keeping its amplitude fixed  \cite{Shchukin2005}.

\section{\label{sec:mg}Quantum state characterization via higher-order moments}

The concept of higher-order moments is connected to the moment generating function \cite{S.M.Barnett1997, Barnett2018} taking for single-mode state the form
\begin{align}
M(\mu) = \sum_{n}(1-\mu)^{n}\rho_{n} = \sum_{m}\frac{{g}^{(m)}}{m!}(-\mu\braket{n})^{m},
\label{eq:M}
\end{align}
which is defined in terms of the variable $\mu$ ($0\le\mu\le2$), mean photon number $\braket{n}$ and the photon statistics $\rho_{n}$. The moment generating function offers a more versatile access to the quantum state characterization and can deliver, for example, parameters such as the photon statistics and all orders of the factorial moments of photon number \cite{Perina2005, Wasilewski2008, Beenakker2001}. Additionally, there is a direct connection to the photon-number parity, delivering the value of the Wigner function, the symmetric-ordered quasi-distribution function  at the origin of the phase space \cite{Cahill1969a, Royer1977, Englert1993}. It can be accessed directly via
\begin{align}
M(\mu = 2) = \sum_{n}(-1)^{n}\rho_{n} = \sum_{m}\frac{{g}^{(m)}}{m!}(-2\braket{n})^{m}.
\end{align}
The moment generating function also allows one to re-express the photon-number parity in terms of the normalized higher-order moments and the loss-corrected mean photon number. The correction for losses is straightforward, since the measured loss-degraded mean photon number is just divided by the efficiency, with which it is being detected \cite{S.M.Barnett1997}. Such a method has experimentally been demonstrated for reconstructing the photon-number parity of heralded single photons generated via PDC emission from a semiconductor waveguide \cite{Laiho2019}. 

Nonetheless, for characterization tasks based on the measurement of higher-order moments, care has to be taken that the summation ofn the right hand side of Eq.~(\ref{eq:M}) converges. If that is the case, the higher-order moments offer expedient means for the direct probing of the Wigner function by reconstructing  the photon-number parity of the states displaced in the phase-space similar to Refs \cite{Banaszek1996, S.Wallentowitz1996, Leibfried1996, P.Bertet2002}. These methods rely on displacing the single-mode state in the phase space with a complex amplitude $\alpha = \textrm{Re}(\alpha) +\imath \textrm{Im}(\alpha)$ corresponding to the transformation $\varrho \rightarrow \hat{D}^{\dagger}(\alpha)\varrho\hat{D}(\alpha) $, in which $\hat{D}(\alpha) $ is the displacement operator in phase space. Experimentally, the displacement operator is easy to realize in an all-optical manner by overlapping the studied state with a coherent displacement beam at an asymmetric beam splitter \cite{DAriano1995, Paris1996}.

\begin{figure*}[!t]
\centering \includegraphics[width = 1.0\textwidth]{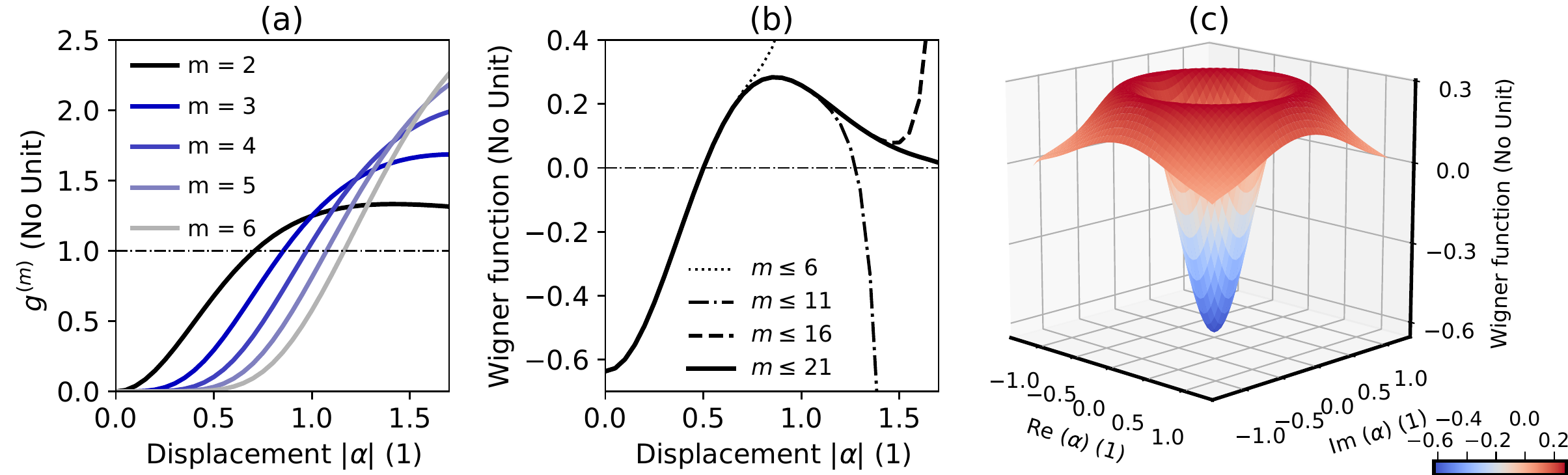}
\caption{\label{fig:gmD}Characteristics of the displaced single-photon state, $\hat{D}(\alpha)\ket{1}$, extracted via higher-order normalized moments. In (a) we illustrate the normalized moments of the displaced single photons in terms of the absolute value of the displacement $|\alpha|$ upto the 6th order given by $g^{(m)}_{\alpha} = \frac{|\alpha|^{2(m-1)[m^{2}+|\alpha|^{2}]}}{(1+|\alpha|^{2})^{m}}$ \cite{Laiho2012,Mahran1986}.  In (b) we show the single photon Wigner function reconstructed by estimating the photon-number parity of the displaced single photon at the displacement $|\alpha|$ via the alternating sum in Eq.~(\ref{eq:wigner}). The summation is truncated upto the 6th (dotted line), 11th (dash-dotted line), 16th (dashed line) and 21st (solid line) moment. 
In (c) we visualize the three dimensional Wigner function corresponding to the solid line in (b).
The mean photon number of the displaced single photon required for the reconstruction is given by $\braket{n}_{\alpha} = 1+|\alpha|^{2}$.
The horizontal dash-dotted lines provide guides for the eye.}
\end{figure*}

The complete Wigner function can be probed via the alternating sum of the displaced state's photon statistics $ \rho_{n}(\alpha)$ via 
\begin{align}
W(\alpha) & = \frac{2}{\pi}\sum_{n}(-1)^{n}\braket{n |\hat{D}^{\dagger}(\alpha)  \varrho \hat{D}(\alpha) |n} \nonumber \\
&
= \frac{2}{\pi}\sum_{n}(-1)^{n}\rho_{n}(\alpha) \nonumber  \\
& = \frac{2}{\pi} \sum_{m}\frac{(-2)^{m}}{m!} {g}^{m}_{\alpha} \braket{n}_{\alpha}^{m}.
\label{eq:wigner}
\end{align}
However, as shown in the last line of Eq.~(\ref{eq:wigner}) it can also easily be extracted via the higher-order moments, in other words, via the normalized $m$-th order moments of the displaced state $g^{(m)}_{\alpha}$ and  its mean photon number $\braket{n}_{\alpha}$.
This definition of the phase-space properties is most suitable, when reconstructing the properties of the Wigner function in discrete variables. 
We note that it is different from the one used for the phase quadratures in Sec.~\ref{sec:cv}, which again is optimally suited for the detection in continous variables. 
In Fig.~\ref{fig:gmD} we theoretically reconstruct the single-photon Wigner function, whose characteristics are phase independent \cite{Nehra2019}. Fig.~\ref{fig:gmD}(a)  shows the characteristics of the normalized higher-order moments of single photons displaced in phase space until the 6th order in terms of the magnitude $|\alpha|$ of the displacement, whereas Fig.~\ref{fig:gmD}(b) presents the values of the alternating sum in Eq.~(\ref{eq:wigner}) by taking into account different amount of moments. The truncation of the summation causes distortions and limits the useful state reconstruction range in the phase space. Clearly, even in an ideal case  without experimental imperfections moments up to $m \approx 21$ are required for reconstructing the interesting range of the single-photon Wigner function depicted wholly in Fig.~\ref{fig:gmD}(c).

\begin{figure*}[!t]
\centering \includegraphics[width = 1.0\textwidth]{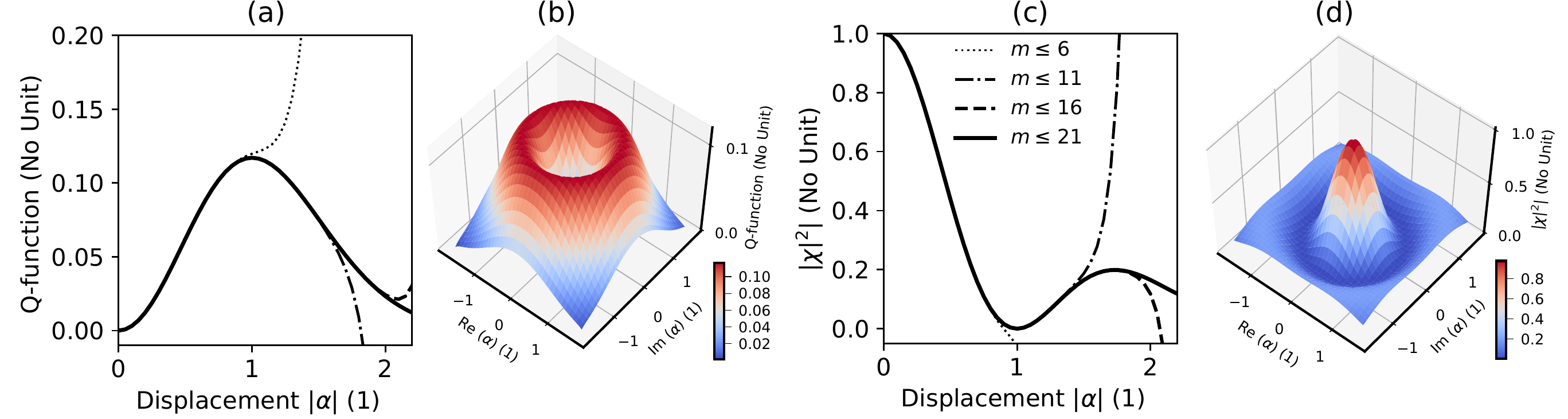}
\caption{\label{fig:p0PDC}  (a) Q-function as well as (b) its three dimensional counterpart in case $m\le21$ and (c) the absolute value squared of the characteristic function of the single-photon state as well as (d) its three dimensional counterpart in case $m\le21$  reconstructed from
the normalized higher-order  moments of the displaced states and their mean photon number. The range of the reconstruction is crucially influenced by the truncation of the summations in Eqs~(\ref{eq:Q}) and (\ref{eq:chi}) upto the 6th (dotted line), 11th (dash-dotted line), 16th (dashed line) and 21st (solid line) moment in (a) and (b), correspondingly. }
\end{figure*}

This way also the Husimi Q-function corresponding to the vacuum contribution of the displaced state and the magnitude of the symmetric-ordered characteristic function has been probed at least for free-propagating single photons \cite{Laiho2012} and for the lowest number states in trapped ions \cite{Ziesel2013, Wolf2019}. The Q-function that can be extracted via the vacuum contribution of the displaced state $\rho_{0}(\alpha)$, can be  very reliably reconstructed, since all the measured orders of the higher-order moments contribute to the summation, which is given by
\begin{align}
Q(\alpha) &=\frac{1}{\pi} \braket{\alpha | \ \varrho \ | \alpha } = \frac{1}{\pi} \braket{0| \hat{D}^{\dagger}(\alpha)  \varrho \hat{D}(\alpha)|0} = \frac{1}{\pi} \rho_{0}(\alpha) 
\nonumber \\
& =\frac{1}{\pi}  \sum_{m} \frac{(-1)^{m}}{m!}g^{(m)}_{\alpha}\braket{n}_{\alpha}^{m}.
\label{eq:Q}
\end{align}

The characteristic function usually cannot be experimentally accessed via photon counting. In the symmetric-order its given $\chi(\alpha) = \textrm{Tr}\{\hat{D}(\alpha)\varrho\}$. With this respect the number states $\ket{n}$ form an exception. 
The absolute value square of their symmetric-ordered characteristic function $\chi(\alpha)_{\ket{n}} = \braket{n | \hat{D}(\alpha) |n}$ is given by
\begin{align}
|\chi(\alpha)|^{2}_{\ket{n}} &= \braket{n|  \hat{D}^{\dagger}(\alpha) |n}\braket{n|\hat{D}(\alpha)|n} = \rho_{n}(\alpha)_{\ket{n
}} \nonumber\\
& =  \sum_{m} \frac{(-1)^{m}}{n! m!} g^{(m+n)}_{\alpha}\braket{n}^{m+n}_{\alpha},
\label{eq:chi}
\end{align}
and corresponds to the probability of the $n$-th photon-number contribution of the displaced state. Clearly only moments higher than or equal to $n$ contribute to the summation.  In Fig.~\ref{fig:p0PDC} we illustrate the reconstruction of both the Q-function and the absolute value squared of the characteristic function for single photons when taking into account different amount of the higher-order normalized moments.  Altogether, the simultaneous access to several characteristic properties of a quantum state offers an in-depth insight to the quantum optical state under study.

\section{\label{sec:concl}Conclusions}

The  characteristics of the photon statistics can be investigated via  photon correlations in a straightforward fashion.
Moreover, the normalized higher-order moments are especially expedient in such state characterization tasks, in which the high optical losses wash out quantum features and prohibit their direct evaluation from the photon-number contributions.
Depending on the temporal coherences of the investigated light, the photon correlations can be investigated in time-dependent or 
time-averaged form. 
While for continuous-wave light usually the time-dependent form provides most detailed information,
the time-averaged form is most suitable for pulsed light. 
Apart from delivering information of the photon-number content of the emitted light, the normalized higher-order moments are routinely employed for accessing the intrinsic properties of the photonic sources.
Additionally, they are feasible for evaluating the state's non-classicality and can even allow for reconstructing the state's phase-space characteristics.
Certainly, care has to be taken when measuring the normalized higher-order moments both in the discrete and continuous variables, such that
enough statistics is taken, which can experimentally turn challenging.
Moreover, the highest reliably resolved moment restricts the accuracy, at which the state characteristics can be extracted affecting also their reliable reconstruction range.
We believe that the photon correlations remain important in quantum optical state classification and characterization tasks and their usage continues evolving along the development of optical detection techniques at the single- and few photon level both with and without true photon-number resolution.

\section*{Acknowledgements}
We thank M. von Helversen for laboratory support, J. Beyer at PTB for caring for the TES detector, and 
A.~E.~Lita and S.~W.~Nam from NIST, USA, for providing the transition-edge-sensor chips.



\begin{thebibliography}{100}
\bibitem{click_det}
The click detectors have the loss-degraded POVMs $\{\hat{\Pi}^{\textrm{no
  click}} =\sum_{n}(1-\eta)^{n}\ket{n}\bra{n} \approx 1- \eta \sum_{n} n
  \ket{n}\bra{n} = 1-\eta \hat{N}$ and $\hat{\Pi}^{\textrm{click}} = 1-
  \hat{\Pi}^{\textrm{no click}} \approx \eta \hat{N} \}$ that deliver at the
  detection efficiency $\eta$ the expectation values of no-click and click
  events, in other words their probabilites, and at low efficiencies ($\eta \ll
  1$) these POVMs can be approximated to be linear with respect to the mean
  photon number. The linear approximation can be guaranteed only if light with
  very low mean photon number impinge on the detector. Thus, the coicidence
  probability in Eq.~(9) is $P_{c} = \braket{\hat{\Pi}^{\textrm{click}}_{1}
  \otimes \hat{\Pi}^{\textrm{click}}_{2}} \approx
  \eta_{1}\eta_{2}\braket{\hat{N}_{1}\hat{N}_{2}}$ and the single click
  probabilities are $P_{D_{1}} \approx \eta_{1}\braket{\hat{N}_{1}}$ and
  $P_{D_{2}} \approx \eta_{2}\braket{\hat{N}_{2}}$.
\bibitem{ph_det}
We utilize the loss-degraded POVMs of $\hat{\Pi}^{m} =
  \sum_{n>m}\left(\begin{array}{c}n\\m\end{array}\right )
  (1-\eta)^{n-m}\eta^{n}\ket{n}\bra{n}$ with $m= 1$ for the true
  photon-number-resolving detection and $\hat{\Pi}^{\textrm{click}} = 1-
  \sum_{n}(1-\eta)^{n}\ket{n}\bra{n}$ for the click detection.
\end{thebibliography}
\end{document}